\documentclass[a4paper,11pt]{JHEP3}
\usepackage{amssymb,amsmath}
\usepackage{axodraw}
\usepackage{graphicx}
\newcommand{\picb}[1]{\;\parbox[c]{45pt}{\begin{picture}(45,30)(0,0)
\SetWidth{1.0}\SetScale{1.0} #1 \end{picture}}\;}
\newcommand{\picc}[1]{\;\parbox[c]{60pt}{\begin{picture}(60,40)(0,0)
\SetWidth{1.0}\SetScale{1.0} #1 \end{picture}}\;}

\title{Classical approximation to quantum cosmological correlations}

\author{Meindert van der Meulen and Jan Smit \\
Institute for Theoretical Physics, University of Amsterdam,\\
Valckenierstraat 65, 1018 XE Amsterdam, The Netherlands.}


\preprint{ITFA-2007-28}

\abstract{ We investigate up to which order quantum effects can be
neglected in calculating cosmological correlation functions after
horizon exit. As a toy model, we study $\phi^3$ theory on a de Sitter
background for a massless minimally coupled scalar field $\phi$. We
find that for tree level and one loop contributions in the quantum
theory, a good classical approximation can be constructed, but for
higher loop corrections this is in general not expected to be
possible. The reason is that loop corrections get non-negligible
contributions from loop momenta with magnitude up to the Hubble scale
$H$, at which scale classical physics is not expected to be a good
approximation to the quantum theory. An explicit calculation of the
one loop correction to the two point function, supports the argument
that contributions from loop momenta of scale $H$ are not
negligible. Generalization of the arguments for the toy model to
derivative interactions and the curvature perturbation leads to the
conclusion that the leading orders of non-Gaussian effects generated
after horizon exit, can be approximated quite well by classical
methods. Furthermore we compare with a theorem by Weinberg. We find
that growing loop corrections after horizon exit are not excluded,
even in single field inflation.}

\begin{document}

\section{Introduction}
The precision of measurements of temperature fluctuations in the
Cosmic Microwave Background radiation has increased enormously in the
recent past and is expected to increase even more in the near
future. From these measurements, statistical properties of the
primordial cosmological perturbations can be deduced. They are found
to have a nearly flat power spectrum, and to be close to
Gaussian. Non-Gaussian effects (see \cite{Bartolo:2004if} for a
review) might be detected in the future and can provide a powerful
tool to discriminate between different inflation models.

There is therefore a large interest in calculating the statistical
properties of the primordial cosmological perturbations for different
inflation models. In these calculations the cosmological perturbations
are often parameterized by the curvature perturbation $\zeta$, which
is the perturbation in scalar curvature on time slices of uniform
density.  This parameterization is convenient because of the property
that, under certain conditions, perturbations in $\zeta$ remain
constant after their wavelengths have grown larger than the horizon
length, i.e. after horizon exit. This has been shown for linear
perturbation theory in \cite{Bardeen:1980kt}, to all orders in $\zeta$
for single field inflation in \cite{Maldacena:2002vr}, and
nonperturbatively for adiabatic perturbations in
\cite{Lyth:2004gb}. The latter two references use a derivative
expansion and therefore assume that effects from wavelengths of the
order of the horizon and shorter, are negligible.

In more complicated models for inflation, e.g. those involving
multiple fields, the curvature perturbation $\zeta$ is not constant
after horizon exit. Therefore evolution after horizon exit might lead
to non-Gaussian effects, which has been investigated in
\cite{Bernardeau:2002jy, Bernardeau:2002jf, Enqvist:2004bk,
Rigopoulos:2004ba, Rigopoulos:2005xx, Rigopoulos:2005ae,
Rigopoulos:2005us, Lyth:2005du, Lyth:2005fi, Zaballa:2006pv,
Vernizzi:2006ve, Allen:2005ye, Lee:2005bb, Kim:2006te, Byrnes:2006vq,
Seery:2006js, Alabidi:2006hg, Seery:2006vu, Battefeld:2006sz,
Battefeld:2007en, Choi:2007su, Yokoyama:2007uu, Byrnes:2007tm}. These
investigations have been done by solving classical equations of
motion, which is assumed to be a good approximation to the quantum
theory, because quantum effects are presumably negligible for
wavelengths much longer than the horizon length (see
\cite{Lyth:2006qz} for a recent argument).

The goal of this paper is to investigate up to which order corrections
to cosmological correlation functions that are generated after horizon
exit, can be calculated reliably using classical physics. For this we
study $\phi^3$ theory on an exact de Sitter background for a massless
minimally coupled scalar field $\phi$, as toy model for the curvature
perturbation $\zeta$ on an inflationary background. We consider
correlation functions with (external) momenta much smaller than the
Hubble scale $H$ in the quantum theory, using the Closed Time Path
(CTP) formalism (also known as in-in formalism, see
e.g. \cite{Calzetta:1986ey, Calzetta:1986cq}), as is also done by
Weinberg in \cite{Weinberg:2005vy,Weinberg:2006ac}. Furthermore we
formulate a classical theory with statistical fluctuations, in such a
way that correlation functions in this theory can easily be compared
with those in the quantum theory.

In short our results for the $\phi^3$ toy model are that the tree
level contributions in the quantum theory can be approximated quite
well by classical physics (which sounds trivial, but we argue that
this is not completely so). However we find that the classical
approximation is not expected to be good in general for loop
corrections in the quantum theory. The reason is that loop integrals
get contributions from momenta with magnitude up to the Hubble scale,
and the contributions from loop momenta at scales around $H$ are in
general not negligible. This is supported by an explicit calculation
of the one loop correction to the two point function, where we use a
small mass as infrared regulator. We argue that the classical
approximation is not expected to be good at scales around $H$, and can
therefore in general not reproduce these contributions. An exception
is the one loop correction, for which we find that the classical
approximation can be saved by choosing a suitable ultraviolet cutoff.

We generalize the arguments for the toy model to derivative
interactions, and apply them to the curvature perturbation
$\zeta$. This leads to the conclusion that non-Gaussian effects
generated after horizon exit in multifield inflation models can, up to
one loop level, be approximated by classical physics. We also compare
with a theorem derived by Weinberg \cite{Weinberg:2005vy}. We find
that it is not excluded that there are corrections to correlation
functions of $\zeta$ that grow after horizon exit, even in single
field inflation.

We remark that the problem we are addressing, is related to, but
different from the problem of the quantum-to-classical transition
\cite{Polarski:1995jg, Kiefer:1998qe, Kiefer:2006je, Martineau:2006ki,
Burgess:2006jn, Prokopec:2006fc}, that deals with the way how quantum
fluctuations acquire classical properties by decoherence, and with the
production of entropy. In this paper we are not investigating how a
quantum system evolves to a classical system; we are considering a
quantum system and a classical system separately from each other and
investigate how well the classical system can reproduce correlation
functions of the quantum system.

In the next section we recall the CTP formalism, using a variation of
the Keldysh-basis, and apply it to $\phi^3$ theory on a de Sitter
background (for other applications of the CTP formalism to interacting
fields in cosmology see \cite{Tsamis:1994ca, Tsamis:1996qm,
Onemli:2002hr, Brunier:2004sb, Onemli:2004mb, Kahya:2006hc,
Prokopec:2002uw, Prokopec:2003qd, Prokopec:2006ue, Boyanovsky:1997cr,
Boyanovsky:2003ui, Boyanovsky:2004gq, Boyanovsky:2004ph,
Boyanovsky:2005sh, Sloth:2006az, Sloth:2006nu, Chaicherdsakul:2006ui,
Bilandzic:2007nb}). Subsequently we analyze contributions to
correlation functions with small external momenta, that are generated
after horizon exit (but still during inflation). In section
\ref{classicaltheory} we show how correlation functions in classical
$\phi^3$ theory on a de Sitter background can be calculated
perturbatively, starting from given initial conditions. The
perturbative contributions are graphically represented in a way that
is similar to the Feynman diagrams of the quantum theory. Next in
section \ref{classicalapproximation} we argue that the classical
theory can approximate the contributions from small internal momenta
in the quantum theory quite well, but that this is in general not the
case for large internal momenta. The one loop correction is an
exception: we show that by choosing a suitable ultraviolet cutoff, the
classical approximation can be good. We generalize our arguments and
conclude in section \ref{discussion}.

In Appendix \ref{appwavefunction} details of the quantization are
given and a comparison is made with finite temperature field
theory. Appendix \ref{appnoextgr} contains an argument on closed
retarded loops. In Appendix \ref{appdiagqcl} diagrams in the classical
and quantum theory are compared. To illustrate the arguments in this
paper, we give in Appendix \ref{app1loopcorr} the detailed
calculations of the one loop correction to the two point function in
the $\phi^3$ toy model, that is generated after horizon exit.

\section{Quantum theory}
\label{quantumtheory}
In this section we set up the quantum theory using the CTP formalism,
and analyze the contributions to correlation functions that are
generated after horizon exit.

The Lagrangian density of $\phi^3$ theory is
\begin{equation}
\mathcal{L}[\phi] = \sqrt{-g} \left( -\frac{1}{2} \partial_{\mu} \phi
\partial^{\mu} \phi - \frac{1}{2} m^2 \phi^2 - \frac{1}{2} \xi R
\phi^2 - \frac{\lambda}{3!}  \phi^3 \right) + \delta \mathcal{L}
\label{lagrangian} 
\end{equation}
where we are using a metric $g_{\mu \nu}$ with signature
$-$$+$$+$$+$. Except when we need the mass $m$ as infrared regulator,
we take $m=0$ and $\xi=0$ to obtain a massless minimally coupled
scalar field. The term $\delta \mathcal{L}$ contains the counterterms:
\begin{equation}
\delta \mathcal{L} = \sqrt{-g} \left( -\delta_1 \phi -\frac{1}{2}
\delta_Z \partial_{\mu} \phi \partial^{\mu} \phi - \frac{1}{2}
\delta_m \phi^2 - \frac{\delta_{\lambda}}{3!} \phi^3 \right) .
\end{equation}
We added a linear counterterm $\delta_1$ to keep $\langle \phi \rangle
= 0$ for all times, at one loop level\footnote{It can be checked that
this is possible by calculating the tadpole diagram, using the $F$ two
point function of equation \eqref{freef} and the infrared and
ultraviolet regulators as discussed later in this paper.}, hence up to
order $\mathcal{O}(\lambda^3)$. The potential can be stabilized by
adding a $\phi^4$ term if desired. We use a spatial momentum cutoff
$\Lambda$ as ultraviolet regulator. Then
\begin{equation}
\delta_m = \frac{\lambda^2}{4 (2\pi)^2} \ln \frac{\Lambda}{\mu} +
\mathcal{O} (\lambda^4), \label{deltas}
\end{equation}
where $\mu$ is a renormalization scale.

\subsection{Closed Time Path formalism on a de Sitter background}

\subsubsection{CTP formalism}
In a system with time-dependent Hamiltonian $H(t)$, that starts in
a state $| {\rm in} \rangle$ at initial time $t_i$, the expectation
value of an operator $Q$ at time $t>t_i$ is given by
\begin{equation}
\langle Q(t) \rangle = \left\langle {\rm in}\left| \left[\bar{\rm T}
  \exp\left( i \int_{t_{\rm in}}^{t} dt' \, H(t') \right) \right] Q
  \left[{\rm T} \exp \left( -i \int_{t_{\rm in}}^t dt'\, H(t') \right)
  \right] \right| {\rm in} \right\rangle ,
\end{equation}
where ${\rm T}$ means a time-ordered product and $\bar{\rm T}$ an
anti-time-ordered product. In the Closed Time Path (CTP) formalism (or
in-in formalism) \cite{Calzetta:1986cq} this expectation value can
also be calculated using path integrals, from the generating
functional
\begin{multline}
Z[J_+,J_-,\rho(t_{\rm in})] = \int \mathcal{D} \phi^+_{\rm in} \mathcal{D}
\phi^-_{\rm in} \langle \phi^+_{\rm in} | \rho(t_{\rm in}) |
\phi^-_{\rm in} \rangle \;
\times \\ \int_{\phi^+_{\rm in}}^{\phi^-_{\rm in}} \mathcal{D} \phi^+
\mathcal{D} \phi^- \; \exp \left[ i \int_{t_{\rm in}}^t dt' \int d^3 x
\left(\mathcal{L}[\phi^+] - \mathcal{L}[\phi^-] + J_+ \phi^+ + J_-
\phi^- \right) \right] . \label{ctpz}
\end{multline}
The path integral on the second line can be written in short-hand
notation as
\begin{equation}
\int \mathcal{D} \phi \; \exp\left[ i \int_{\mathcal{C}} dt' \int d^3 x \,
  \left( \mathcal{L}[\phi] + J \phi \right) \right],
\end{equation}
where $\mathcal{C}$ is the so-called Schwinger-Keldysh contour which
runs from $t_{\rm in}$ to $t$ and back. The field $\phi$ and source
$J$ are split up in $\phi^+$, $J_+$ on the first part of this contour,
and $\phi^-$, $J_-$ on the second part, with the condition $\phi^+(t)
= \phi^-(t)$. The integration along the contour $\mathcal{C}$ explains
the name Closed Time Path formalism. The path integral on the first
line of equation \eqref{ctpz} imposes that at the initial time $t_{\rm
in}$, the state of the system is given by the density matrix
$\rho(t_{\rm in})$.  Expectation values are then obtained by variation
of the sources $J_+$ and $J_-$:
\begin{multline}
\langle {\rm \bar{T}} \left(\phi(x_1) \ldots \phi(x_n) \right) {\rm T}
\left( \phi(x_{n+1}) \ldots \phi(x_{n+m}) \right) \rangle = \\
\frac{\delta^{n+m} Z[J_+,J_-,\rho(t_{\rm in})]}{\delta J_- (x_1)
\cdots \delta J_- (x_{n}) \; \delta J_+(x_{n+1}) \cdots \delta
J_+(x_{n+m})} \Bigg|_{J_+,J_-=0},
\end{multline}
where the times $x_j^0$ are smaller than or equal to the time $t$ used
in definition \eqref{ctpz}. 

When calculating these correlation functions perturbatively, we need
to know the free two point functions with all four possible time
orderings: 
\begin{align}
G^{-+}(x,y) &=  i \langle \phi(x) \phi(y) \rangle^{(0)},
\label{gmp} \\ 
G^{+-}(x,y) &=  i \langle \phi(y) \phi(x) \rangle^{(0)},
\label{gpm} \\ 
G^{++}(x,y) &=  i \langle {\rm T} \phi(x) \phi(y) \rangle^{(0)} =
\theta(x_0-y_0) G^{-+}(x,y) + \theta(y_0-x_0) G^{+-}(x,y), \\
G^{--}(x,y) &=  i \langle {\rm \bar{T}} \phi(x) \phi(y) \rangle^{(0)} =
\theta(x_0-y_0) 
G^{+-}(x,y) + \theta(y_0-x_0) G^{-+}(x,y),
\end{align}
where the superscript $(0)$ denotes the free field correlation
functions. They obey the identity
\begin{equation}
G^{++}(x,y) + G^{--}(x,y) = G^{-+}(x,y) + G^{+-}(x,y), \label{gid}
\end{equation}
and they can be put together in a matrix:
\begin{equation}
\mathbf{G}(x,y) = \left( \begin{array}{cc}
G^{++}(x,y) & G^{+-}(x,y) \\
G^{-+}(x,y) & G^{--}(x,y) \end{array} \right).
\end{equation}
Note that the two point functions depend on the initial conditions via
the dependence on $\rho(t_i)$ of the generating functional
\eqref{ctpz}.

In the context of the classical approximation it is useful to
transform the $\phi^+$ and $\phi^-$ fields to a different basis, which
is a variation of the Keldysh basis (see also \cite{Aarts:1997kp}):
\begin{equation}
\left( \begin{array}{c} \phi^{(1)} \\ \phi^{(2)} \end{array}\right)  = 
\left( \begin{array}{c} (\phi^+ + \phi^-)/2 \\ \phi^+ - \phi^-
\end{array} \right) =
\mathbf{R} \left( \begin{array}{c} \phi^+ \\ \phi^- 
\end{array} \right), \hspace{.6cm} \mbox{with} \hspace{.6cm} 
\mathbf{R} = \left(
\begin{array}{cc} 1/2 & 1/2 \\ 1 & -1 \end{array} \right) .
\end{equation}
The Lagrangian density $\mathcal{L}[\phi^+]-\mathcal{L}[\phi^-]$
transforms to
\begin{equation}
\mathcal{L}[\phi^{(1)}, \phi^{(2)}] = \sqrt{-g} \left( -
\partial_{\mu} \phi^{(1)} \partial^{\mu} \phi^{(2)} - (m^2 + \xi R)
\phi^{(1)} \phi^{(2)} - \frac{\lambda}{3!} \left( 3 (\phi^{(1)})^2
\phi^{(2)} + \frac{1}{4} (\phi^{(2)})^3 \right)
\right). \label{lagph12} 
\end{equation}
The free two point functions in this basis can easily be obtained by
the transformation
\begin{equation}
\mathbf{G}_K(x,y) = \mathbf{R} \mathbf{G}(x,y) \mathbf{R}^T = \left(
\begin{array}{cc} i F (x,y) & G^R (x,y) \\ G^A(x,y) & 0 \end{array}
\right),
\end{equation}
with
\begin{align}
F (x,y) & = -\frac{i}{2} \left( G^{-+}(x,y) + G^{+-}(x,y) \right),
\label{qfp} \\
G^R(x,y) &= G^{++}(x,y) - G^{+-}(x,y) = \theta(x_0-y_0) \left(
G^{-+}(x,y) - G^{+-}(x,y) \right), \label{qrp}\\
G^A(x,y) &= G^{++}(x,y) - G^{-+}(x,y) = \theta(y_0-x_0) \left(
G^{+-}(x,y) - G^{-+}(x,y) \right), \label{qap}
\end{align}
where we have used identity \eqref{gid}. They obey the equations
\begin{align}
\left(\Box_x + m^2 + \xi R(x) \right) F(x,y) &= 0, \\
\left(\Box_x + m^2 + \xi R(x) \right) G^{R,A}(x,y) & = \frac{\delta^4
  (x-y)}{\sqrt{-g(x)}}, \label{greqdef}
\end{align}
with 
\begin{equation}
\Box_x = \frac{1}{\sqrt{-g(x)}} \, \partial_{\mu} \left(\sqrt{-g(x)} \,
g^{\mu \nu} (x) \partial_{\nu} \right).
\end{equation}
The $G^R$ and $G^A$ two point functions are often called the retarded
and advanced propagators. Note that $G^A(x,y) = G^R(y,x)$.

\subsubsection{Feynman rules on a de Sitter background}
\label{feynrules}
The metric of the de Sitter background is
\begin{equation}
ds^2 = -dt^2 + a^2(t)\, d\mathbf{x}^2, \label{dsmetric}
\end{equation}
where $a(t)$ is the FRW scale factor. The Hubble rate is $H=\dot
a/a$. In de Sitter space the scale factor is $a(t)=a_0 \exp(Ht)$. We
will use conformal time $\tau= -\int_t^{\infty} dt'/a(t')$, which runs
from $-\infty$ to $0$. The scale factor in conformal time is $a(\tau)
= -1/H\tau$.

As initial state $\rho(\tau_{\rm in})$ we take the adiabatic or
Bunch-Davies vacuum, for $\tau_{\rm in}\to -\infty$. One expects that
other choices will give the same results because this state is an
attractor state \cite{Anderson:2000wx, Anderson:2005hi}. The free
field operator with this initial state is given in equation
\eqref{modedecom} in Appendix \ref{appwavefunction} and it can be used
to derive the free two point functions \eqref{qfp}-\eqref{qap}:
\begin{align}
F(k,\tau_1,\tau_2) &= \frac{H^2}{2 k^3} \left[ (1+k^2 \tau_1 \tau_2)
  \cos k(\tau_1 - \tau_2) + k(\tau_1-\tau_2) \sin k (\tau_1 - \tau_2)
  \right], \label{freef} \\ 
G^R(k,\tau_1,\tau_2) &= \theta(\tau_1-\tau_2) \frac{H^2}{
   k^3} \left[(1+k^2 \tau_1 \tau_2) \sin k(\tau_1 - \tau_2) - k
  (\tau_1 - \tau_2) \cos k(\tau_1 - \tau_2) \right], \label{freegr}
\end{align}
and $G^A(k,\tau_1,\tau_2) = G^R(k,\tau_2,\tau_1)$, and where the two
point functions depend only on the length of the spatial momentum $k =
|\mathbf{k}|$. 
Representing the $\phi^{(1)}$ field with a full line and the
$\phi^{(2)}$ field with a dashed line, the Feynman rules for the two
point functions, the vertices and the counterterm are\footnote{These
  Feynman rules should not be confused with the graphical
  representation developed in \cite{Musso:2006pt}.}
\begin{align}
\picb{
\Line(2,15)(44,15)
\Text(4,24)[lt]{$\tau_1$}
\Text(43,24)[rt]{$\tau_2$}
} 
\hspace{.0cm} & = F(k,\tau_1,\tau_2), \\
\picb{
\Line(2,15)(22,15)
\DashLine(22,15)(44,15){3}
\Text(4,24)[lt]{$\tau_1$}
\Text(43,24)[rt]{$\tau_2$}
} 
\hspace{.0cm} & = -i G^R(k,\tau_1,\tau_2) = -i
G^A(k,\tau_2,\tau_1), 
\label{propsfeyn} \\
\picb{
\DashLine(4,15)(22,15){3}
\Text(8,17)[lb]{$\tau_1$}
\Line(22,15)(40,28)
\Text(25,24)[lb]{$\tau_2$}
\Line(22,15)(40,2)
\Text(25,7)[lt]{$\tau_3$}
} & = -i \lambda \, a^4(\tau_1) \delta(\tau_1-\tau_2) \delta(\tau_1-\tau_3),
\label{cvertex} \\  
\picb{
\DashLine(4,15)(22,15){3}
\Text(8,17)[lb]{$\tau_1$}
\DashLine(22,15)(40,28){3}
\Text(25,24)[lb]{$\tau_2$}
\DashLine(22,15)(40,2){3}
\Text(25,7)[lt]{$\tau_3$}
} & = -\frac{i \lambda}{4} a^4(\tau_1) \delta(\tau_1-\tau_2) \delta(\tau_1
- \tau_3), 
\label{qvertex} \\
\picb{
\DashLine(4,15)(19.9,15){2}
\Line(24.1,15)(40,15)
\CArc(22,15)(3,0,360)
\Line(19.9,17.1)(24.1,12.9)
\Line(19.9,12.9)(24.1,17.1)
\Text(10,18)[bc]{$\tau_1$}
\Text(34,18)[bc]{$\tau_2$}
} \hspace{.02\textwidth} 
& = -i a^4(\tau_1) \delta (\tau_1 - \tau_2) \, \delta_m . \label{ctfeyn}
\end{align}
When a two point function is attached to a vertex, the corresponding
time has to be integrated over. A closed loop corresponds with an
integral over spatial momentum $\int d^3p/(2 \pi)^3$.

\subsubsection{Example: equal time two point function}
\label{extwopoint}
As an example to which we will return repeatedly, we consider the
equal time two point function up to one loop level:
\begin{equation}
\int d^3 x \; e^{-i \mathbf{k} \cdot \mathbf{x}}
\langle\phi(\tau,\mathbf{x}) \phi(\tau,\mathbf{0}) \rangle .
\label{def2point} 
\end{equation}
The tree level contribution is given by
\begin{equation}
\picb{
\Line(2,15)(44,15)
\Text(4,24)[lt]{$\tau$}
\Text(43,24)[rt]{$\tau$}
} 
\hspace{1.5cm} F (k,\tau,\tau).
\end{equation}
There is no contribution with the $G^R$ two point function because
that vanishes for equal times. At one loop level there are
contributions
\begin{equation}
\picc{
\Line(2,20)(12,20)
\Text(4,29)[lt]{$\tau$}
\DashLine(12,20)(22,20){2}
\CArc(32,20)(10,90,180)
\DashCArc(32,20)(10,0,90){2}
\CArc(32,20)(10,180,360)
\Line(42,20)(60,20)
\Text(58,29)[rt]{$\tau$}
\Text(15,42)[lt]{A}
} \hspace{.4cm}
\picc{
\Line(2,20)(12,20)
\Text(4,29)[lt]{$\tau$}
\DashLine(12,20)(22,20){2}
\CArc(32,20)(10,0,360)
\DashLine(42,20)(52,20){2}
\Line(52,20)(60,20)
\Text(58,29)[rt]{$\tau$}
\Text(15,42)[lt]{B}
} \hspace{.4cm}
\picc{
\Line(2,20)(12,20)
\Text(4,29)[lt]{$\tau$}
\DashLine(12,20)(22,20){2}
\CArc(32,20)(10,90,270)
\DashCArc(32,20)(10,-90,90){2}
\DashLine(42,20)(52,20){2}
\Line(52,20)(60,20)
\Text(58,29)[rt]{$\tau$}
\Text(15,42)[lt]{C}
} \hspace{.4cm}
\picc{
\Line(2,20)(12,20)
\Text(4,29)[lt]{$\tau$}
\DashLine(12,20)(22,20){2}
\CArc(25,20)(3,0,360)
\Line(22.9,22.1)(27.1,17.9)
\Line(22.9,17.9)(27.1,22.1)
\Line(28,20)(46,20)
\Text(44,29)[rt]{$\tau$}
\Text(15,42)[lt]{D}
}, \label{1loopdiagrams}
\end{equation}
where diagrams A, C and D have also mirror versions which correspond
to interchanging the endpoints. There is also the diagram 
\begin{equation}
\picc{
\Line(2,20)(22,20)
\Text(4,29)[lt]{$\tau$}
\CArc(32,20)(10,90,180)
\DashCArc(32,20)(10,180,270){2}
\CArc(32,20)(10,-90,0)
\DashCArc(32,20)(10,0,90){2}
\Line(42,20)(60,20)
\Text(58,29)[rt]{$\tau$}
}, \label{vanishingdiagram}
\end{equation}
but it vanishes because of the $\theta$-functions in the $G^R$ two
point functions. This is an example of the general fact that diagrams
with no external $G^R$ two point functions vanish, as explained in
Appendix \ref{appnoextgr}. Furthermore there are diagrams with dashed
lines at the endpoints, but these diagrams vanish also because of
$\theta$-functions. Diagrams with tadpoles are canceled by the linear
counterterm $\delta_1$.

Using the Feynman rules, the diagrams can be translated to
integrals. For example diagram A translates to
\begin{multline}
\picc{
\Line(2,20)(12,20)
\Text(4,29)[lt]{$\tau$}
\DashLine(12,20)(22,20){2}
\CArc(32,20)(10,90,180)
\DashCArc(32,20)(10,0,90){2}
\CArc(32,20)(10,180,360)
\Line(42,20)(60,20)
\Text(58,29)[rt]{$\tau$}
\Text(15,42)[lt]{A}
} \hspace{.3cm}
=(-i)^2 (-i \lambda)^2 \int_{\tau_{\rm in}}^{\tau} d\tau_1\; a^4(\tau_1)\;
\int_{\tau_{\rm in}}^{\tau} d\tau_2\; a^4(\tau_2)\; G^R(k,\tau,\tau_1)
F(k,\tau,\tau_2) \; \times \\ \int\frac{d^3p}{(2\pi)^3}
G^R(|\mathbf{k} + 
\mathbf{p}|,\tau_1, \tau_2) F(p,\tau_1,\tau_2). \label{diagA} 
\end{multline}
The symmetry factor is $1$ because the propagators in the loop are
different (diagrams B and C have symmetry factor $1/2$). The momentum
integral is both infrared and ultraviolet divergent. The ultraviolet
divergence is the same as the one that occurs in Minkowski space and
is canceled by the counterterm in diagram D. The infrared divergence
has to be regularized, e.g. by giving $\phi$ a small mass or by taking
space to be finite.

In Appendix \ref{app1loopcorr} the one loop correction
\eqref{1loopdiagrams} is calculated explicitly, using an initial time
$\tau_H$ with $|k\tau_H|<1$.

\subsection{Late times}
\label{latetimes}

\subsubsection{Cosmological correlation functions}
In this paper we consider cosmological correlation functions, by which
we mean equal time correlation functions
\begin{equation}
\int d^3 x_1 \ldots d^3 x_{r} \, e^{-i \mathbf{k}_1 \cdot
  \mathbf{x}_1 - \ldots - i \mathbf{k}_{r} \cdot \mathbf{x}_{r}} \langle
\phi(\tau,\mathbf{x}_1) \ldots \phi(\tau,\mathbf{x}_{r}) \phi(\tau,
  \mathbf{0}) \rangle, 
  \label{etcorrfunc}  
\end{equation}
where the time $\tau$ is late, i.e. is well after horizon exit with
respect to the spatial momenta $k_i$, which can be expressed as $|k_i
\tau| \ll 1$. Similarly, early times are times for which $|k_i \tau|
\gg 1$. 

We constrain the analysis further by only considering contributions to
these cosmological correlation functions, that are generated after
horizon exit. For this we introduce a split in time at $\tau_H$, a few
($N_H$) e-folds after horizon exit, such that $|k\tau_H| = \exp(-N_H)
\ll 1$. Correlation functions at $\tau_H$ have accumulated
contributions from earlier times, of which we only keep the free field
contributions. Then we use these correlation functions as initial
conditions for the evolution after $\tau_H$. In practice this means
that we use the Feynman rules as described in the previous subsection,
with the only difference that we take $\tau_H$ as initial time. This
procedure is not completely correct because the neglected
contributions generated before $\tau_H$ are of the same order in the
coupling constant as the contributions generated after $\tau_H$. But
it does not change the qualitative behaviour of the contributions
generated after $\tau_H$, and neglecting the contributions from before
$\tau_H$ simplifies the calculations significantly. We comment further
on this point in section \ref{earlytimecontrs}.

Different contributions to cosmological correlation functions
\eqref{etcorrfunc} depend in different ways on the time $\tau$. Those
that are proportional to positive powers of $\tau$, are dominated by
their values at the initial time $\tau_H$. These contributions are
negligible if $\tau_H$ is taken sufficiently long after horizon
exit. Contributions that are proportional to a non-positive power of
$\tau$ can grow after horizon exit and will therefore dominate. In
\cite{Weinberg:2005vy,Weinberg:2006ac} it is shown that these
contributions do not grow faster than powers of $\ln(-H\tau)$, so
negative powers of $\tau$ do not occur. In this paper we call
contributions that are proportional to $\tau^0$ (including powers of
$\ln(-H\tau)$) late time contributions. In this subsection we analyze
the dependence on $\tau$ of the different contributions by counting
powers of $\tau$.

Contributions can contain integrals over spatial internal (loop)
momenta, $p$, which can be arbitrarily large. We have found that the
power counting goes quite differently for small internal momenta
(smaller than the Hubble scale $H$, i.e. $|p\tau| \ll 1$), than for
large internal momenta (of the order of the Hubble scale and larger,
i.e. $|p \tau| \gtrsim 1$). Therefore we analyze first the case that
all internal momenta are small, and consider then arbitrary
(amputated) 1PI diagrams for which all the internal momenta are large
and the external momenta are small. These 1PI diagrams can be treated
as effective (non-local) couplings in the analysis for small internal
momenta, and in this way our analysis covers the whole range of
internal momenta. After this general analysis, we compare our results
with the specific case of the one loop correction to the two point
function.

\subsubsection{Small internal momenta}
\label{sims}
When the internal momenta are small, the expressions for the free two
point functions $F$ and $G^R$ \eqref{freef}, \eqref{freegr} can be
expanded in $k\tau_i$:
\begin{align}
F(k,\tau_1, \tau_2) =& \frac{H^2}{2 k^3} [1 + \mathcal{O}(k^2
  \tau_i^2) ] , \label{expanf} \\ 
G^R(k,\tau_1, \tau_2) =& \theta(\tau_1-\tau_2) \frac{H^2}{3 k^3} [ k^3
  (\tau_1^3 - \tau_2^3) + \mathcal{O}(k^5 \tau_i^5) ], \label{expangr}
\end{align}
where $k^2 \tau_i^2$ indicate all possible combinations $k^2
\tau_1^2$, $k^2 \tau_1 \tau_2$ and $k^2 \tau_2^2$, and similarly for
$k^5 \tau_i^5$. Using the lowest order of these expansions, it is easy
to count the powers of $\tau_i$ of the contribution of an arbitrary
Feynman diagram. The $F$ two point function does not contribute any
factor of $\tau_i$, and the $G^R$ two point function gives a factor of
$\tau_i^3$. Furthermore, a vertex contributes a factor $a^4(\tau_i)
\propto \tau_i^{-4}$, and an integral $\int d\tau_i \propto \tau_i$,
so effectively it contributes a factor $\tau_i^{-3}$.

We can divide the Feynman diagrams into two classes: diagrams that
contain only vertices with one dashed line (diagrams A, B and D in the
example of the two point function \eqref{1loopdiagrams}), and diagrams
that contain one or more vertices with three dashed lines (diagram C
in \eqref{1loopdiagrams}). Because each dashed line is attached to a
$G^R$ two point function, diagrams of the first class have an equal
number of vertices as $G^R$ two point functions. Each vertex
contributes a factor $\tau_i^{-3}$ and each $G^R$ two point function a
factor of $\tau_i^3$ and therefore diagrams from the first class are
proportional to $\tau_i^0$. Because these diagrams can contain integrals
like $\int d\tau_i/\tau_i$, they can be proportional to powers of $\ln
(-\tau_i)$, which are largest for the upper limit of the time
integrals, i.e. $\tau$. 

Diagrams from the second class have more $G^R$ two point functions
than vertices, and are therefore proportional to positive powers of
$\tau_i$. For example a diagram with one vertex with three dashed
lines has two more $G^R$ two point functions than if this vertex would
have had one dashed line, and is therefore suppressed by a factor of
$|k\tau_i|^6$. The contributions from diagrams of the second class are
largest for the lower limit of the time integrals, i.e. for $\tau_H$.
Therefore the contributions of diagrams from this class are suppressed
with respect to contributions of the first class by a factor of
$|k\tau_H|^6=\exp(-6 N_H)$, and they do not have the growing factors
of $\ln (\tau/\tau_H)$.

The power counting of $\tau_i$ that is done here, is similar to the
power counting of the scale factor $a$ in the derivation of the
theorem in \cite{Weinberg:2005vy}. The expansion of the two point
functions \eqref{expanf} and \eqref{expangr} can be compared with the
asymptotic expansions of the wavefunctions for late times in
\cite{Weinberg:2005vy}. The difference is that here we use
power counting to differentiate between growing and vanishing
contributions to correlation functions for a specific ($\phi^3$)
interaction, whereas in \cite{Weinberg:2005vy} it was used to
differentiate between interactions leading to different late time
behaviour. We return to this in section \ref{othertheories}.

\subsubsection{Large internal momenta}
\label{lims}
We consider an arbitrary amputated 1PI diagram, which has small
external momenta, and where we integrate the internal momenta starting
at a scale $M$ somewhat smaller than $H$. We use a cutoff $\Lambda$ as
ultraviolet regulator. Because both $M$ and $\Lambda$ are physical
scales and not comoving scales, the limits of the momentum integrals
are time dependent: $M a(\tau_i)$ and $\Lambda a(\tau_i)$, where
$\tau_i$ corresponds with the time of one of the vertices. We take for
this time the earliest time that occurs in the loop, because that
corresponds with the smallest cutoff.\footnote{If an ultraviolet
divergence is local, it occurs only if the times of the vertices in
the loop are equal (i.e. they are proportional to
$\delta(\tau_{j_1}-\tau_{j_2})$ for all times $\tau_{j_i}$); then it
does not matter which time one chooses.}  In this way, the momentum
integrals do also contribute time dependencies, which we also have to
take into account.

Suppose that our arbitrary amputated 1PI diagram has $E$ external
lines and $N$ vertices, each with $V$ legs. Then there are $P$
internal lines and $L$ loops with
\begin{align}
P & = \frac{1}{2} \left( N V - E \right), \label{propsnumber} \\
L &= \frac{1}{2} \left(N V - E \right) - N + 1 . \label{loopsnumber}
\end{align}
Furthermore from equations \eqref{freef} and \eqref{freegr} we see
that each internal two point function contributes factors proportional
to
\begin{equation}
\frac{(p_i \tau_{j_1})^{n_{\beta}}}{p_i^3} \, e^{\pm i p_i \tau_{j_2}},
\end{equation}
where $n_{\beta}=0,1,2$. Each vertex gives $\int d\tau_j/\tau_j^4$,
and each loop gives an integral $\int d^3 p_i$. We ignore powers of
the external momenta because the internal momenta are much larger, $p
\gg k$. Then we can count the powers of $p$ and $\tau$ of the diagram:
\begin{equation}
 p_i^{(-3+n) P+3 L-l}\, \tau_j^{-3 N+ n P-l} \, \bigg|^{\Lambda
   a(\tau_j)}_{M a(\tau_j)} \to
\left(\frac{\Lambda}{H} \right)^{\frac{n}{2}(N V-E) - 3 N - l+3} \,
\tau_j^{-3}, \label{lmscaling} 
\end{equation}
where $n$ is the sum of the $n_{\beta}$, and where $l$ is a
non-negative integer that represents the fact that $p$ integrals can
also lead to factors of $1/\tau$ instead of an extra factor $p$ (see
e.g. the integrals \eqref{lmcontr1}-\eqref{lmcontr9} in the example
calculation of Appendix \ref{app1loopcorr}). We ignore the
contribution from the lower limit $M a(\tau_j)$, because the full result
cannot depend on the split in the integrals.\footnote{This is
confirmed for the specific case considered in Appendix
\ref{app1loopcorr}.}

Apparently the power of $\tau_j$ is independent of the details of the
calculation: contributions from large internal momenta to 1PI diagrams
are always proportional to $\tau_j^{-3}$. 

We can now compare the 1PI diagram, seen as an effective coupling,
with a tree level coupling. They are both proportional to
$\tau_j^{-3}$, but they differ in the possible numbers of (external)
dashed lines. The 1PI diagrams can have any number of dashed lines,
instead of one or three dashed lines for the tree level
coupling. However, as shown in Appendix \ref{appnoextgr}, it turns out
that 1PI diagrams with no external dashed lines vanish. Hence the
non-vanishing 1PI diagrams can have one or more external dashed lines.

The 1PI diagrams, being effective couplings for small momenta, can be
put in the analysis for small internal momenta of section
\ref{sims}. The power counting argument of that section shows that
only those effective couplings with one external dashed line can lead
to contributions proportional to $\tau^0$. Effective couplings with
more external dashed lines lead to contributions that are suppressed
by (at least) a factor $|k\tau_H|^3$.

The power of $\Lambda/H$ in equation \eqref{lmscaling} does depend on
the details of the calculation. When it is non-negative, it can cause
an ultraviolet divergence. Some divergent terms are proportional to
$\delta(\tau_{j_1}- \tau_{j_2})$ for all the times $\tau_{j_i}$ that
occur in the loop. These divergences are the usual local divergences
and are canceled by counterterms. Non-local divergent terms can also
occur (an example of this is given below), as a consequence of the way
in which the total correction is split up in contributions from
individual diagrams. They must cancel between different contributions,
in order to make the total result finite.

It is interesting to consider the errors that occur when the
ultraviolet regulator $\Lambda$ is not taken to infinity but kept
finite. From equation \eqref{lmscaling} it appears that, after the
counterterms have been taken into account, the errors will be
proportional to positive powers of $H/\Lambda$. Hence if $\Lambda$ is
taken to be smaller than $H$, large errors occur, but if $\Lambda$ is
taken to be larger than $H$ the errors are suppressed. Clearly,
internal momenta of the order magnitude of the Hubble scale $H$ still
contribute to the correlation functions, even though $H$ is much
larger than the external momenta $k/a(\tau)$. This is a feature of
quantum field theory in de Sitter space that is different from what
one would expect from field theory in flat space. In the latter case,
from the point of view of effective field theories, one expects only
contributions from internal momenta of the order of magnitude of the
external momenta. Contributions from higher scales are said to
decouple. In de Sitter space, scales decouple only when they are
larger than the Hubble scale $H$.

\subsubsection{Example: Late time contributions to equal time two point
  function at one loop} 
The general analysis of this subsection can be checked in the example
of the late time contributions to the one loop correction to the two
point function \eqref{1loopdiagrams}, as calculated in Appendix
\ref{app1loopcorr}. In this calculation the momenta are split up
between small and large at a comoving scale $M_{\rm cm}$, which obeys
$|M_{\rm cm} \tau| \ll 1$ and $M_{\rm cm} > k$.

\paragraph{Small internal momenta.}
The results for small internal momenta of diagrams A and B are given
in equations \eqref{contrsm} and \eqref{resdiagblow},
respectively. Diagram C does not give any late time
contribution. After attaching the external lines, the dominant terms
can be found in equations \eqref{evad} and \eqref{evbc} and are
proportional to
\begin{equation}
\frac{\lambda^2}{\epsilon} \ln^2 \frac{\tau}{\tau_H} \qquad
\mbox{with} \qquad \epsilon = \frac{m^2}{3 H^2}, \qquad \mbox{or}
\qquad \lambda^2 \ln^3 \frac{\tau}{\tau_H},
\label{quickestgrowthsm} 
\end{equation}
depending on the values of $\epsilon$ and $\ln(\tau/\tau_H)$, where
$\epsilon$ is the infrared regulator. In this calculation we have used
an expansion that is valid for $|\epsilon \ln(- k \tau)| < 1$, and
therefore this calculation is only valid for a limited amount of
time. When $|\epsilon \ln(-k \tau)|$ approaches $1$, the term on the
right in \eqref{quickestgrowthsm} becomes of comparable magnitude to
the term on the left.

Two powers of the logarithm $\ln (\tau/\tau_H)$ come from the two time
integrals corresponding with the two vertices. The extra factor of
$\ln(\tau/\tau_H)$ in the term on the right in
\eqref{quickestgrowthsm}, is the consequence of the momentum
integration, and was observed earlier in a similar calculation in
\cite{Boyanovsky:2004ph}, where also a small mass was used as infrared
regulator.

\paragraph{Large internal momenta.}
For large momenta the result for the amputated diagrams are given in
\eqref{resdiagalarge} for diagram A, \eqref{uvdiagb} and
\eqref{ltclmb} for diagram B and \eqref{uvdiagc} for diagram
C. Diagram A has a local ultraviolet divergence that is canceled by
the counterterm, diagram D. Diagrams B and C have divergent terms that
are non-local, and that cancel each other. The finite remainder is
suppressed for late times. The contribution \eqref{ltclmb} from
diagram B only removes the dependence on the scale $M_{\rm
cl}$. Therefore only diagram A leads to late time contributions, in
agreement with the result above that late time contributions can only
come from 1PI diagrams with one external dashed line.

The term from diagram A that grows quickest for large internal
momenta, after attaching the external lines, can be found in equation
\eqref{evad} and is proportional to
\begin{equation}
\lambda^2 \ln^3 \frac{\tau}{\tau_H},
\end{equation}
which is comparable to the term on the right in
\eqref{quickestgrowthsm} for the small internal momenta.

\paragraph{Complete result for the one loop correction.}
The complete result of the late time contributions to the one loop
correction is two times equation \eqref{evad} added to equation
\eqref{evbc}, which gives
\begin{align}
\frac{\lambda^2}{36 (2\pi)^2 k^3} \bigg\{ & \frac{7}{9 \epsilon} +
\frac{392}{27} - \frac{7}{3} \gamma - \frac{17}{18} \pi^2 -\frac{4}{3}
\ln 2 - 4 \zeta(3)
- \ln \frac{2 \mu}{H} + \frac{4}{9} \ln(- k \tau_H) + \nonumber \\
& \hspace{-.1\textwidth} \bigg(\frac{2}{\epsilon}+15-\frac{17}{3}
\gamma - \frac{2}{3} \pi^2 
-\frac{8}{3} \ln 2- 3\ln \frac{2 \mu}{H} + \frac{8}{3} \ln(- k \tau_H)
\bigg) \ln \frac{\tau}{\tau_H} + \nonumber \\ 
& \hspace{-.1\textwidth} \bigg(\frac{2}{\epsilon} + \frac{22}{3} - 2
\gamma -2 \ln 2+ 4 \ln(-k \tau_H) \bigg) \ln^2 
\frac{\tau}{\tau_H} +\frac{8}{3} \ln^3 \frac{\tau}{\tau_H} +
\mathcal{O} (\frac{\tau}{\tau_H} ) + \mathcal{O}(\epsilon)
\bigg\}. \label{endresult} 
\end{align}

Note that a consequence of the growing behaviour of loop corrections
is that the theory become nonperturbative if one waits long enough.

\section{Classical theory}
\label{classicaltheory}
In this section we consider classical $\phi^3$ theory for a massless
minimally coupled field $\phi$. The evolution of classical fields on a
de Sitter background is governed by the equation of motion (which can
be derived from the Lagrangian density \eqref{lagrangian})
\begin{equation}
\partial_{\tau}^2 \phi(x)+ 2 H a(\tau) \partial_{\tau} \phi(x) -
\nabla^2 \phi(x) + a^2 (\tau) \frac{\lambda}{2!} \phi^2(x) =
0, \label{cleom} 
\end{equation}
where we use $x=(\tau,\mathbf{x})$ with $\tau$ conformal time. Initial
conditions have to be imposed at an initial time $\tau_{\rm in}$. We
focus on the calculation of equal time correlation functions
\begin{equation}
\langle \phi(\tau, \mathbf{x}_1) \ldots \phi(\tau, \mathbf{x}_n)
\rangle_{\rm cl}, \label{clcorrfunc} 
\end{equation}
where the subscript ``cl'' denotes a correlation function in the
classical theory. In this section we show how to calculate these
correlation functions in a way that is similar to the interaction
picture in quantum field theory: first we calculate the free field
correlation functions starting from the initial conditions and using
the free equations of motion, and then we calculate perturbative
corrections, expressed in terms of these free field correlation
functions. In \cite{Aarts:1996qi,Aarts:1997kp} this method was used in
the context of thermal field theory. Furthermore, we show that the
contributions to the correlation functions can be represented
graphically in a way that is similar to Feynman diagrams.

\subsection{Perturbative calculation of correlation functions}
We assume that at the initial time $\tau_{\rm in}$, initial conditions
are given for the correlation functions
\begin{equation}
\langle \phi(\tau_{\rm in},\mathbf{x}_1) \ldots \phi(\tau_{\rm in},
\mathbf{x}_n) \rangle_{\rm cl},
\end{equation}
and first order time derivatives of these correlation functions. In
the free field case ($\lambda = 0$), the initial conditions can be
evolved in time using the free field equations of motion. Then one
obtains the free field correlation functions
\begin{equation}
\langle \phi_0(\tau_1,\mathbf{x}_1) \ldots \phi_0(\tau_n, \mathbf{x}_n)
\rangle_{\rm cl}, \label{freefieldcorrs}
\end{equation}
where the subscript ``0'' denotes the free field solutions, and where
the times $\tau_1,\ldots, \tau_n$ do not have to be equal. 

To calculate perturbative corrections to the correlation functions, we
first solve the classical equation of motion for $\phi(x)$
perturbatively. The first order correction is
\begin{equation}
\phi_1(x) = -\frac{\lambda}{2!} \int d^4 y\, a^4(y_0)
G^R(x,y) \phi_0^2(y),
\end{equation}
where $y_0$ denotes conformal time, and where the retarded propagator
$G^R(x,y)$ is the solution of
\begin{equation}
\frac{1}{a^2 (x_0)}\left(\partial_{x_0}^2 + 2 H a(x_0) \partial_{x_0}
- \nabla^2 
\right) G^R(x,y) = \frac{\delta^4(x-y)}{a^4(x_0)},
\end{equation}
where we used de Sitter metric \eqref{dsmetric}. This equation is the
same as \eqref{greqdef}, so that the retarded propagator in the
classical theory is equal to the one in quantum theory, given in
equation \eqref{freegr} after a spatial Fourier transform. Higher
order perturbative corrections to the solution of the equation of
motion are obtained by
\begin{equation}
\phi_i (x) = -\frac{\lambda}{2!} \int d^4 y\, a^4(y_0) \,
G^R(x,y) \sum_{j=0}^{i-1} \phi_j(y) \phi_{i-j-1}(y). \label{pertsol}
\end{equation}
By iteration the $i$-th order solution can be expressed in terms of
the zeroth order solution $\phi_0 (x)$. The full perturbative solution
of the equation of motion \eqref{cleom} is the sum
\begin{equation}
\phi(x) = \sum_{i} \phi_i (x). \label{clsolsum}
\end{equation}

Perturbative corrections to the correlation function
\eqref{clcorrfunc} are obtained by replacing the $\phi(x_i)$'s in
\eqref{clcorrfunc} by the perturbative solution \eqref{clsolsum}, and
ordering the terms according to the total powers of $\lambda$:
\begin{equation}
\langle \phi(\tau, \mathbf{x}_1) \ldots \phi(\tau, \mathbf{x}_n)
\rangle_{\rm cl} = \sum_r \, \langle 
\phi(\tau,\mathbf{x}_1) \ldots \phi(\tau,\mathbf{x}_n) \rangle_{\rm
  cl}^r, 
\end{equation}
with
\begin{equation}
\langle
\phi(\tau,\mathbf{x}_1) \ldots \phi(\tau,\mathbf{x}_n) \rangle_{\rm cl}^r = 
\sum_{i_1+ \ldots + i_n = r} \langle \phi_{i_1}
(\tau,\mathbf{x}_1) \ldots \phi_{i_r} (\tau, \mathbf{x}_n)
\rangle_{\rm cl}. \label{clcorrfunr} 
\end{equation}
When the $\phi_{i_j}$'s are completely expressed in terms of free
field solutions $\phi_0(x)$, the corrections to the correlation
function \eqref{clcorrfunr} are expressed in terms of free field
correlation functions, which we have obtained from the free equations
of motion and the initial conditions in \eqref{freefieldcorrs}.

\subsection{Graphical representation}
\label{graphrep}
When we choose the initial conditions to be Gaussian, it is possible
to represent the contributions on the right hand side of equation
\eqref{clcorrfunr} graphically in a way that is similar to the Feynman
diagrams of the quantum theory. In the free field case, Gaussian
initial conditions evolve to Gaussian free field correlation
functions. Therefore the free field correlation functions are
completely determined by the two point function, which we call
suggestively $F_{\rm cl}$: 
\begin{equation}
F_{\rm cl}(x_1, x_2) = \langle \phi_0 (x_1) \phi_0 (x_2) \rangle_{\rm
  cl}. \label{deffcl}
\end{equation}

We assign graphical rules analogously to the quantum case:
\begin{align}
\picb{
\Line(2,15)(22,15)
\DashLine(22,15)(44,15){3}
\Text(4,24)[lt]{$x$}
\Text(43,24)[rt]{$y$}
} 
\hspace{.5cm}& = -i G^R(x,y), \\
\picb{
\Line(2,15)(44,15)
\Text(4,24)[lt]{$x$}
\Text(43,24)[rt]{$y$}
} 
\hspace{.5cm} & = F_{\rm cl}(x,y), \label{fclgraph} \\
\picb{
\DashLine(4,15)(22,15){3}
\Line(22,15)(40,28)
\Line(22,15)(40,2)
} \hspace{.05\textwidth} & = \frac{-i \lambda}{2} \int d^4 y \, a^4
(y_0), 
\end{align}
and furthermore
\begin{equation}
\picb{
\Line(2,15)(16,15)
\Line(14,13)(18,17)
\Line(14,17)(18,13)
\Text(2,20)[cb]{$x$}
} \hspace{.5cm} \phi_0 (x). \label{clendpoint}
\end{equation}

The $r$-th order corrections on the right-hand side of equation
\eqref{clcorrfunr} can be constructed graphically in two steps. First
the $\phi_i$'s of equation \eqref{pertsol} are represented by tree
graphs where the endpoint $x$ and the $i$ vertices are connected to
each other by $G^R$ propagators. The remaining free legs of the
vertices are occupied by the $\phi_0$'s of \eqref{clendpoint}. For
example the second order solution $\phi_2(x)$ can be represented by
\begin{multline}
\picb{
\Line(2,6)(10,6)
\DashLine(10,6)(18,6){2}
\Text(2,11)[cb]{$x$}
\Line(18,6)(26,1)
\Line(18,6)(26,11)
\DashLine(26,11)(34,16){2}
\Line(34,16)(42,11)
\Line(34,16)(42,21)
\Line(40,9)(44,13)
\Line(40,13)(44,9)
\Line(40,19)(44,23)
\Line(40,23)(44,19)
\Line(24,-1)(28,3)
\Line(24,3)(28,-1)
} \hspace{.5cm}
\phi_2(x) = 2 \int d^4 y\;
a^4(y_0) \; (-i) G^R(x,y) \frac{-i \lambda}{2!} \phi_0(y) \; \times \\
\int d^4 z \; 
a^4(z_0) \; (-i) G^R(y,z) \frac{-i \lambda}{2!} \phi_0^2 (z),
\end{multline}
where the factor $2$ comes from two equal contributions.

In the second step the tree graphs representing the $\phi_{i_j}$ in
\eqref{clcorrfunr} are glued together at the crosses in all possible
ways. When two crosses are glued together, a full line is created
representing the free two point function \eqref{fclgraph}. Consider
for example the contribution $\langle \phi_2 (x_1) \phi_0 (x_2)
\rangle_{\rm cl}$ to the second order two point function $\langle
\phi(x_1) \phi(x_2) \rangle_{\rm cl}^{(2)}$. The tree graphs
representing $\phi_2(x_1)$ and $\phi_0(x_2)$ can be glued together in
two ways:
\begin{multline}
\picc{
\Line(2,20)(12,20)
\DashLine(12,20)(22,20){2}
\Text(2,25)[cb]{$x_1$}
\CArc(32,20)(10,90,180)
\DashCArc(32,20)(10,0,90){2}
\CArc(32,20)(10,180,360)
\Line(42,20)(58,20)
\Text(58,25)[cb]{$x_2$}
} \hspace{.8cm}
(-i)^2 (-i\lambda)^2 \int d^4 y\;
a^4(y_0) \; G^R(x_1,y) \; \times \\ \int d^4 z \;
a^4(z_0) \; G^R(y,z) \, F_{\rm cl}(y,z)\, F_{\rm cl}(z,x_2),
\label{clth2pt1} 
\end{multline}
(where an extra factor 2 comes from two ways of contracting the
$\phi_0$'s), and
\begin{multline}
\picc{
\Line(2,2)(15,2)
\Text(2,5)[cb]{$x_1$}
\DashLine(15,2)(30,2){2}
\Line(30,2)(30,10)
\DashLine(30,10)(30,22){2}
\CArc(30,30)(8,0,360)
\Line(30,2)(58,2)
\Text(58,5)[cb]{$x_2$}
} \hspace{.8cm}
\frac{(-i)^2 (-i\lambda)^2}{2} \int d^4 y\, a^4 (y_0) \, G^R(x_1,y) \,
F_{\rm cl} (y,x_2)  \times \\
\int d^4 z \, a^4(z_0) \, G^R(y,z) \, F_{\rm
  cl}(z,z). \label{clthtadpole} 
\end{multline}
The former diagram is equal to diagram A in the quantum theory
(equation \eqref{1loopdiagrams}, after a spatial Fourier transform) if
$F_{\rm cl}(x_1,x_2)$ equals the $F$ two point function in equation
\eqref{freef}.

Both diagrams \eqref{clth2pt1} and \eqref{clthtadpole} can be
divergent, depending on $F_{\rm cl}$. If the divergences are local
they can be canceled by adding counterterms. In fact diagram
\eqref{clthtadpole} contains a tadpole diagram, which is automatically
local and can be canceled completely by a linear counterterm.

Similarly one can construct the contribution $\langle \phi_0(x_1)
\phi_2(x_2) \rangle_{\rm cl}$, which is equal to the mirror version of
diagram A in \eqref{1loopdiagrams}. Finally there is $\langle
\phi_1(x_1) \phi_1(x_2) \rangle_{\rm cl}$, which is equal to diagram B
in the quantum theory. It is not possible to obtain diagram C in the
classical theory.

Note that the resulting classical diagrams can have loops. This
illustrates that loop corrections occur not only in the quantum
theory, but also in the classical theory. These diagrams do not
vanish, because there are statistical fluctuations.

\section{Classical approximation}
\label{classicalapproximation}
In section \ref{quantumtheory} we have investigated the late time
behaviour of the quantum theory, and in section \ref{classicaltheory}
we have set up the classical theory. The graphical representation of
the classical perturbative corrections as described in section
\ref{graphrep} suggests that the classical theory reproduces exactly
the diagrams of the quantum theory with only vertices with one dashed
line. In Appendix \ref{appdiagqcl} a precise argument is given that
shows that this is indeed the case. Hence if we choose the initial
conditions of the classical approximation such that the classical free
field two point function $F_{\rm cl}$ is equal to the quantum $F$ two
point function \eqref{freef}, the classical approximation reproduces
the contributions of the quantum theory coming from these diagrams.

In the one loop correction to the two point function
\eqref{1loopdiagrams}, this means that the classical approximation is
given by diagrams A, B and D (counterterms are still necessary in the
classical approximation).

In this section we investigate how good the classical theory is as a
classical approximation to the quantum theory for late times.

\subsection{Small internal momenta}
\label{clapproxsims}
As argued in section \ref{sims} for small internal momenta, the
diagrams with only vertices with one dashed line give exactly the
contributions in the quantum theory that are proportional to
$\tau^0$. The other diagrams, that have vertices with three dashed
lines and are not in the classical approximation, give contributions
that are suppressed by $|k\tau_H|^6 = \exp(-6 N_H)$, because each
vertex with three dashed lines leads to two more retarded propagators
compared to a vertex with one dashed line. These contributions do not
grow after horizon exit. Therefore, for small internal momenta the
classical approximation is good up to errors that are suppressed by a
factor of $\exp(-6 N_H)$ with respect to the late time contributions.

This is confirmed in the example of the one loop correction to the two
point function. Here the late time contributions from small internal
momenta are completely coming from diagrams A and B and these diagrams
do indeed occur in the classical approximation.

The internal momenta in tree diagrams are always small. Therefore the
tree level contributions in the quantum theory can well be
approximated by a classical approximation, if $\tau_H$ is chosen
sufficiently long after horizon exit. Note that this is not trivial:
the quantum theory contains tree diagrams with vertices with three
dashed lines that do not occur in the classical approximation.

\subsection{Large internal momenta}
\label{clthlims}
In section \ref{lims} we have seen that in the quantum theory, loop
corrections get late time contributions from internal momenta up to
the Hubble scale $H$, from 1PI diagrams with one external dashed
line. This set of diagrams is not the same as the diagrams of the
classical approximation. Namely, the 1PI diagrams with one external
dashed line can contain vertices with three dashed lines, but these
diagrams do not occur in the classical approximation. Therefore, the
classical approximation misses late time contributions from large
internal momenta. This could have been expected: the classical
approximation is not supposed to be good for physics at scales around
$H$. 

For one loop corrections, the classical approximation can be saved: it
turns out that the classical approximation does not miss any late time
contributions, because at one loop level there are no 1PI diagrams
having both one external dashed line and a vertex with three dashed
lines. However, another problem arises: because the classical
approximation has fewer diagrams than the quantum theory, not all
ultraviolet divergences are canceled. In the next subsection, we treat
these new ultraviolet divergences by introducing a cutoff.

In the two point function, this becomes apparent by the fact that the
classical approximation does not contain diagram C. As a consequence,
the ultraviolet divergence of diagram B is not canceled. As mentioned
above, this ultraviolet divergence is not local, and can therefore not
be canceled by a counterterm.

\subsection{Classical approximation at one loop}
\label{clapproxoneloop}
In order to deal with the ultraviolet divergences in the classical
approximation at one loop, we introduce a cutoff at a physical scale
$\Lambda$ (we use a physical scale and not a comoving scale, such that
we can use the arguments of section \ref{lims}). In the following we
estimate the magnitude of the error that is induced by this cutoff, in
one loop diagrams.

First of all there are the ultraviolet divergences that made the
cutoff necessary. From equation \eqref{lmscaling} we see that they are
proportional to $(\Lambda /H)^{n_1}$ with $n_1 \geq 0$ (the case
$n_1=0$ can give a factor $\ln (\Lambda/H)$). As argued above, these
divergences occur (at least at one loop level) only in 1PI diagrams
with two or more external dashed lines, and are therefore, according
to the argument in section \ref{lims}, suppressed by a factor
$\exp(-3N_H)$ or a positive power of this factor. Thus errors coming
from the divergent terms have an extra factor of $\exp(-3 N_H)
(\Lambda/H)^{n_1}$ with respect to late time contributions from large
internal momenta.

This can be checked for the example in Appendix \ref{app1loopcorr}:
the term of the amputated version of diagram B that causes a linear
divergence in the classical approximation is given in
\eqref{uvdiagb}. Together with the external $G^R$ two point functions,
the divergent term has an extra factor $|k \tau_H|^3 (\Lambda/H)$ with
respect to non-vanishing late time contributions, e.g. equation
\eqref{ltclmb}.

Secondly there are the errors from terms that are proportional to
inverse powers of $\Lambda$ and that would vanish if the cutoff would
be sent to infinity. For 1PI diagrams with one external dashed line,
these errors can be proportional to $\tau^0$ and give late time
contributions. These errors therefore have only an extra factor of
$(H/\Lambda)^{n_2}$ with $n_2>0$, with respect to other late time
contributions from large internal momenta.

The total error thus has scales like
\begin{equation}
c_1 \, e^{-3 N_H} \left( \frac{\Lambda}{H} \right)^{n_1} +  c_2 \left(
\frac{H}{\Lambda} \right)^{n_2}, \label{errorestimate}
\end{equation}
with respect to other late time contributions, where $c_i$ are
constants of order $\mathcal{O}(1)$. To make this factor considerably
smaller than one, $\tau_H$ should be chosen long enough after horizon
exit and the cutoff $\Lambda$ should be chosen considerably larger
than the Hubble scale $H$; e.g. if $n_1=n_2=1$ and $c_1=c_2$,
$\Lambda$ should be chosen of the order of $\exp(3N_H/2) H$.

Instead of introducing the cutoff by hand in the momentum integrals,
one can also remove the ultraviolet divergences by changing the
ultraviolet behaviour of the initial conditions. For example one can
put a cutoff in the initial conditions:
\begin{equation}
\int d^3 x \, e^{-i \mathbf{k} \cdot \mathbf{x}} \, \langle \phi(\tau_H,
  \mathbf{x}) \phi(\tau_H, \mathbf{0}) \rangle_{\rm 
  cl} = \frac{H^2}{2 k^3} (1+k^2 \tau_H^2) 
\theta(\Lambda_{\rm cm} -k),
\end{equation}
where $\Lambda_{\rm cm}$ is now a cutoff in comoving momentum. The
physical scale $\Lambda$ that corresponds with $\Lambda_{\rm cm}$ is
$|\Lambda_{\rm cm}/a(\tau)|$, thus the error estimate
\eqref{errorestimate} becomes now
\begin{equation}
c_1 \, e^{-3 N_H} \left| \Lambda_{\rm cm} \tau \right|^{n_1} + c_2
\left| \frac{1}{\Lambda_{\rm cm} \tau} \right|^{n_2}.
\end{equation}
Another possibility is to choose initial conditions as
\begin{equation}
\int d^3 x \, e^{-i \mathbf{k} \cdot \mathbf{x}} \, \langle \phi(\tau_H,
  \mathbf{x}) \phi(\tau_H, \mathbf{0}) \rangle_{\rm 
  cl} = \frac{H^2}{2 k^3} \frac{1+k^2 \tau_H^2}{1+k^2 \tau_c^2},
\end{equation}
where there is not a hard cutoff, but the loop diagrams are made
finite (except for the tadpole diagram, which does not cause problems
because it can be canceled by a local counterterm). The time $\tau_c$
acts like an inverse (soft) cutoff in comoving momenta, so that it
should be taken small enough to make $\tau/\tau_c \gg 1$ and large
enough to keep $|k\tau_H|^3 (\tau/\tau_c)^n$ small.

\subsection{Example}
In Appendix \ref{app1loopcorr} we have calculated the one loop
correction to the two point function in the quantum theory. As an
example, we have computed numerically the error that comes from using
a finite cutoff for a specific set of parameters, as a function of the
cutoff. This is done for diagram A by taking the upper limit in
equation \eqref{verylongeq}, adding the counterterm (diagram D), and
subtracting the term that remains finite in the limit $\Lambda \to
\infty$ (i.e. the second line of \eqref{resdiagalarge}). Then the
external lines are attached and the times $\tau_1$ and $\tau_2$ are
integrated numerically. For diagram B the upper limit in equation
\eqref{verylongeq2} is used, and there are no counterterms. The
results are divided by the full correction \eqref{endresult}, and are
plotted, separately for diagrams A and D, and for diagram B, in figure
\ref{fig:errors}. Both errors are much smaller than the full
correction \eqref{endresult}. The error from diagrams A and D is
clearly decreasing for increasing cutoff. The error from diagram B is
suppressed, but increasing linearly with the cutoff. These results
agree with the arguments given in this section.  
\FIGURE{
\includegraphics[width=.6\textwidth]{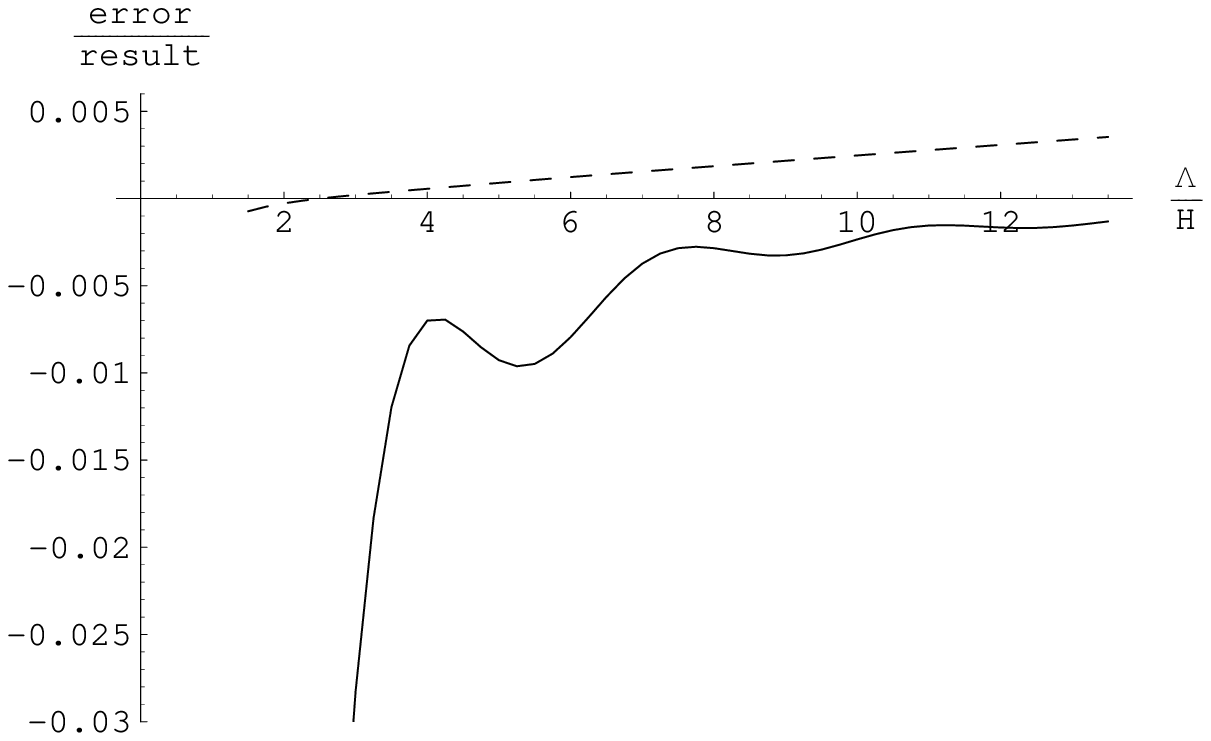}
\caption{Plot of the numerically calculated error with respect to the
  complete result \eqref{endresult}, for diagrams A and D (full) and
  diagram B (dashed), versus the cutoff $\Lambda/H$. We have used
  $k\tau_H=-0.4$, $k\tau = -0.03$, $\epsilon=0.1$, and $2 \mu=H$.}
\label{fig:errors}
}

\section{Discussion and conclusions}
\label{discussion}

\subsection{Early time contributions}
\label{earlytimecontrs}
Both in the quantum theory and in the classical theory we have
neglected early time contributions, i.e. contributions from times
before $\tau_H$. In both cases they can be included by imposing
initial conditions at $\tau_H$, which can be obtained by calculating
equal time correlation functions at $\tau_H$ in the quantum theory
with initial time $\tau_{\rm in} \to -\infty$.

In the quantum theory the initial conditions can be represented by
non-local $n$-point vertex functions that act only at the initial time
$\tau_H$, as explained for example in \cite{Calzetta:1986cq}. These
vertex functions can have any number of dashed and full lines. They
can give by themselves constant late time contributions, and can also
occur in diagrams that give growing late time contributions. These
extra ingredients make the arguments and calculations more
complicated, but do not change them qualitatively. For example the
calculation of the one loop correction to the two point function in
Appendix \ref{app1loopcorr} will have extra contributions from early
times, of order $\lambda^2$ but without growing factors
$\ln(\tau/\tau_H)$.

In the classical theory a similar thing can be done. A practical
problem is that not all the vertex functions can be represented in the
classical theory, because they can have any number of dashed lines,
while in the classical theory only vertices with one dashed line can
be represented. Instead the initial conditions at $\tau_H$ can be
imposed by adjusting the free field correlation functions. They are
then in general non-Gaussian and cannot be represented only by the
free field two point function \eqref{deffcl}. As a consequence, the
calculations become more complicated, but not qualitatively different,
as is the case in the quantum theory.

\subsection{Generalization to derivative interactions}
\label{othertheories}
Throughout this paper we have used $\phi^3$ theory as a toy model. For
other interactions the analysis of this paper can be adjusted, which
is straightforward for $\phi^n$ interactions, but less so for
derivative interactions, which are of particular interest for
cosmology.

Let us first consider contributions from small internal momenta. A
spatial derivative leads to an inverse power of the scale factor, or
equivalently to a factor of $\tau$. As can be seen from the power
counting argument in section \ref{sims}, this extra factor of $\tau$
suppresses late time contributions and prevents that any factors of
$\ln(\tau/\tau_H)$ occur.

For a temporal derivative the situation is more complicated. If a time
derivative $\partial_{t_1}=-H\tau_1 \partial_{\tau_1}$ acts on an $F$
two point function, which can be expanded as \eqref{expanf}, the
constant term vanishes, so the result is proportional to $\tau_i^2$
and as a consequence, late time contributions are suppressed. But if a
time derivative acts on a $G^R$ two point function, which can be
expanded as \eqref{expangr}, the result is still proportional to
$\tau_i^3$, and there is no suppression of late time contributions
yet. Only when there is also a time derivative acting on the other
time argument $\partial_{t_2}$, the result becomes proportional to
$\tau_i^5$ and late time contributions are suppressed.

Hence when an interaction has one time derivative, e.g. $\dot\phi
\phi^2$, there are still late time contributions: if at all vertices
(with one dashed line) the time derivatives act at the dashed line,
no extra factors of $\tau_i$ appear and late time contributions are
not suppressed. But when an interaction has two time derivatives,
e.g. $\dot\phi^2 \phi$, there is no way to avoid suppression of the
late time contributions: in a diagram with $N$ vertices (all of this
type and with one dashed line), there are $N$ $G^R$ two point
functions and $2 N$ time derivatives, so that there must be at least
one $G^R$ two point function with time derivatives on both sides, or
an $F$ two point function with a time derivative.

For large internal momenta, we can reconsider the power counting
argument in section \ref{lims}. A derivative (spatial or temporal)
leads in \eqref{lmscaling} to either an additional factor $p\tau$, or
to no additional factors. Hence derivatives only change the value of
the integer $l$, and do not change the arguments of section
\ref{lims}. Apparently late time contributions from large internal
momenta are not necessarily suppressed by derivative interactions.

In \cite{Weinberg:2005vy} these matters are treated in a slightly
different way. There the time integrals are performed first, and after
that the momentum integrals, which is a different order than employed
in this paper. For fixed external and internal momenta, a theorem is
derived that shows that if the interactions obey certain conditions,
the time integrals converge for $\tau \to 0$. For the wavefunctions an
asymptotic expansion is used, valid for late times (or equivalently
small momenta).

Because in this theorem $\tau \to 0$ is taken, relative to which all
fixed momenta are small, and because of the use of the asymptotic
expansion, this theorem can be compared with our findings above for
small internal momenta. They are indeed in agreement with the
conditions of the theorem.

\subsection{Comparison to stochastic approach}
It is interesting to compare the classical theory as described in
section \ref{classicaltheory}, to the stochastic approach
\cite{Starobinsky:1986fx, Starobinsky:1994bd}. In this approach the
field $\phi$ is also considered to be a classical field with
statistical fluctuations after horizon exit. The difference with the
classical theory of section \ref{classicaltheory} is that the
fluctuations are not imposed as initial conditions at a fixed initial
time, but are put into the system stochastically at wavenumber
$H$ at all times.

The stochastic approach has in the recent literature been used for
different purposes, e.g. for calculating non-Gaussianities
perturbatively in multifield inflation \cite{Rigopoulos:2004ba,
  Rigopoulos:2005xx, Rigopoulos:2005ae, Rigopoulos:2005us}, and for
investigating nonperturbative behaviour in de Sitter space that occurs
at very late times \cite{Woodard:2005cw, Tsamis:2005hd}, when the
factors $\ln(- \tau)$ have grown so large that they overcome the
suppression by small coupling constants. In the latter case it has
been argued \cite{Woodard:2005cw, Tsamis:2005hd} that the stochastic
approach can reproduce the terms with the largest power of $\ln(-
\tau)$ at each order in the coupling constant (leading log
approximation).

Below we first discuss the case of only a massless minimally coupled
scalar field with nonderivative interactions, as in section
\ref{classicaltheory}. Then we make some remarks on theories with
derivative interactions and with other fields than massless minimally
coupled scalars.

\subsubsection{Massless minimally coupled scalar with nonderivative interactions}
The stochastic approach does not use a mass as infrared regulator, but
uses a finite lower limit for the momentum integrals: the classical
field $\phi$ is defined to contain only modes with comoving wavenumber
$k>H$. Physically this corresponds to considering only a finite patch
of de Sitter space, the size of which increases exponentially by the
expansion. So to compare with the classical theory of section
\ref{classicaltheory}, we need to reformulate the latter using this
infrared regulator.

Apart from the different infrared regulator, the stochastic approach
makes two additional approximations with respect to the classical
theory of section \ref{classicaltheory}. First the classical field is
assumed not to contain modes with wavenumber $k>Ha$; hence the
momentum integrals have an upper limit $Ha$. Second, the wavefunction
of the free scalar field \eqref{masslessmodefunctions}, that is used to
characterize the stochastic fluctuations, is approximated by its
leading term for late times: $\phi_{k,1}(\tau) \to i \frac{H}{\sqrt{2
    k^3}}$. In the classical theory of section \ref{classicaltheory}
this is equivalent to taking only the leading term of the expansion of
the $F$ two point function, as is done in equation
\eqref{expanf}. Also the retarded propagator $G^R$ is approximated by
its leading term, as in equation \eqref{expangr}.

We now compare the stochastic approach with the classical theory of
section \ref{classicaltheory} for the one loop correction of the two
point function in $\phi^3$ theory, in particular the term with the
largest power of $\ln(- \tau)$. For the classical theory we can use
the calculation in Appendix \ref{app1loopcorr}; we only adjust the
infrared regulator. This means that in the calculations for small
internal momenta we put $\epsilon \to 0$, and use a lower limit $H$
for the momentum integrals. For diagram A this changes equation
\eqref{contrsm} to
\begin{equation}
\frac{i \lambda^2 \theta(\tau_1-\tau_2)}{6 (2 \pi)^2 H^4 (\tau_1
  \tau_2)^4} \left( 2 (\tau_1^3- \tau_2^3) \ln \frac{M_{\rm cm}}{H}
\right). 
\end{equation} 
The calculation for large internal momenta is unchanged and gives a
finite contribution after adding the counterterm of diagram D.
Attaching the external lines and performing the time integrals gives
for the term with the largest power of $\ln(-\tau)$
\begin{equation}
\frac{\lambda^2}{36 (2 \pi)^2 k^3} \bigg\{-\frac{1}{3} \ln^3
\frac{\tau}{\tau_H} \bigg\}. \label{fsdiagAsims}
\end{equation}
For the small internal momenta of diagram B we use the integral
\begin{equation}
\int_H^{M_{\rm cm}} \frac{dp}{p^2} \int_{|p-k|}^{p+k} \frac{dp'}{
  p'^2} = \frac{1}{k^2} \bigg( \ln \frac{k^2}{H^2} + \ln \frac{M_{\rm cm}-k}{
  M_{\rm cm}+k} + \frac{2k}{M_{\rm cm}} - 2 \bigg),
\end{equation}
so that the analog of \eqref{resdiagblow} becomes 
\begin{equation}
\frac{-\lambda^2}{4 (2 \pi)^2 k^3 H^4 (\tau_1 \tau_2)^4 } \bigg( \ln
\frac{k^2}{H^2} + \ln \frac{M_{\rm cm}-k}{M_{\rm cm} +k} +
\frac{2k}{M_{\rm cm}} - 2 \bigg) . \label{fsdiagBsims}
\end{equation}
Again the integral for the large internal momenta remains unchanged
(we use a cutoff as described in section \ref{clapproxoneloop}), and
attaching the external lines and performing the time integrals gives a
leading logarithmic term of $\ln^2(\tau/\tau_H)$. Apparently there is
no contribution to the $\ln^3 \tau/\tau_H$ term from diagram B when we
use this infrared regularization.

We can calculate the same quantity in the stochastic approach by using
stochastic sources. However we will not do this, but instead repeat
the calculation of above, using the additional assumptions of the
stochastic approach (i.e. taking the leading order approximations for
the propagators \eqref{expanf}, \eqref{expangr}, and the upper limit
$k<H a$ in the momentum integral). We expect that this does not make a
difference for the result of the largest power of $\ln(-\tau)$ and
therefore we interpret the result of this calculation as the result of
the stochastic approach.

With the approximations of the stochastic approach, the calculation of
diagram A reduces to the calculation for small internal momenta done
above, with only the upper limit changed from $M_{\rm cm}$ to $H
a(\tau_2)$. For the amputated version of diagram A the result is then
\begin{equation}
  \frac{i \lambda^2 \theta(\tau_1-\tau_2)}{6 (2 \pi)^2 H^4 (\tau_1
    \tau_2)^4} \left( 2 (\tau_1^3- \tau_2^3)\ln \frac{-1}{H \tau_2}
  \right), 
\end{equation}
and after attaching the external lines and performing the time
integrals, the result for the largest power of $\ln (-\tau)$ is the
same as in our formulation of the classical theory,
\eqref{fsdiagAsims}. For the amputated version of diagram B the result
is
\begin{equation}
\frac{-\lambda^2}{4 (2 \pi)^2 k^3 H^4 (\tau_1 \tau_2)^4 } \bigg( \ln
\frac{k^2}{H^2} + \ln \frac{H a(\tau_2)-k}{H a(\tau_2)+k} +
\frac{2k}{H a(\tau_2)} - 2 \bigg),
\end{equation}
which, similarly to the contribution \eqref{fsdiagBsims} for our
formulation of the classical theory, does not lead to $\ln^3
\tau/\tau_H$ terms.

In this calculation the stochastic approach reproduces the same
leading logarithmic term (but not the same subleading logarithmic
terms) as the classical theory in our formulation, using a lower
momentum limit as infrared regulator. We remark that if we would have
split the momentum integrals at a physical scale $M_{\rm cm} a$
instead of the comoving scale $M_{\rm cm}$ in the calculation of the
classical theory, {\it all} the contributions to the leading
logarithmic term would have come from small internal momenta. It is
reasonable to expect that this remains true for higher
orders. Moreover the approximations for the propagators in the
stochastic approach are then also the same as we made in the
calculation for the small internal momenta. Therefore we expect that
the stochastic approach will give the largest logarithmic term at each
order in the coupling, consistent with the arguments in
\cite{Woodard:2005cw, Tsamis:2005hd}.

Because of the used approximations, the stochastic approach has fewer
problems with the ultraviolet than our formulation of the classical
theory (see sections \ref{clthlims} and \ref{clapproxoneloop}). The
drawback of using these approximations is that even at one loop order,
only the leading logarithmic term can be obtained, whereas in our
formulation of the classical theory, also the subleading logarithmic
terms can be obtained at one loop order, as explained in section
\ref{clapproxoneloop}.

\subsubsection{Derivative interactions and other fields}
\label{stochapprdevs}
As argued in section \ref{othertheories}, derivative interactions
typically lead to positive powers of $\tau$ and therefore to
suppression of late time contributions from small internal
momenta. There can also be interactions with fields that are not
massless minimally coupled scalar fields. The wavefunctions of these
fields are proportional to a positive power of $\tau$ for late times,
instead of to $\tau^0$, as is the case for massless minimally coupled
scalar fields. Therefore the leading terms of the expansions of the
$F$ two point functions of these fields are also proportional to a
positive power of $\tau$. If a diagram contains such an $F$ two point
function, this diagram cannot lead to late time contributions from
small internal momenta.

However, even if there are no late time contributions from small
internal momenta, it is still possible that there are late time
contributions from large internal momenta. The stochastic approach
makes approximations that are not valid for large internal
momenta. Therefore the stochastic approach can have problems with
reproducing the largest logarithmic terms correctly in these cases. In
\cite{Miao:2006pn} the stochastic approach is applied to the theory of
a massless minimally coupled scalar field interacting with a massless
fermion. Here the problem that there are no contributions from small
internal momenta is circumvented by integrating out the fermion, and
considering the effective theory of the scalar field.

\subsection{Application to the curvature perturbation}
The motivation for this work comes from cosmological perturbations
generated during a period of inflation. As mentioned in the
introduction, a suitable parameterization for these cosmological
perturbations is the curvature perturbation $\zeta$, which typically
has interactions involving derivatives. It is of interest to know late
time contributions of correlation functions of $\zeta$, and whether
these can be approximated by a classical approximation.

For small internal momenta, one can derive the interaction terms for
the various degrees of freedom in a specific model of inflation, and
then use the conditions of the theorem of \cite{Weinberg:2005vy} to
decide whether these interactions can lead to late time
contributions. For single field inflation (possibly together with
$\mathcal{N}$ free massless scalar fields) it is shown in
\cite{Weinberg:2005vy} that the interactions do obey the conditions,
and therefore do not give late time contributions to all orders. This
can be compared with \cite{Lyth:2004gb}, where it is argued, using
classical physics, that $\zeta$ is conserved after horizon exit. This
argument is only valid for adiabatic perturbations, which applies to
single field inflation, and for small internal momenta, because the
large internal momenta are removed by a smoothing procedure. Indeed,
we found in section \ref{clapproxsims} that for small internal momenta
the quantum theory can be approximated quite well by classical
physics.

For inflation models involving more fields, there are typically
interactions that do not obey the conditions of the theorem in
\cite{Weinberg:2005vy}, and therefore can lead to late time
contributions. Correspondingly, the perturbations are not adiabatic in
these models, hence the argument of \cite{Lyth:2004gb} does not apply
and it is no surprise that $\zeta$ is not conserved after horizon
exit.

For large internal momenta the situation is different. Since
derivatives do not necessarily suppress late time contributions, it is
possible that loop corrections lead to late time contributions, even
for single field inflation. Moreover, a classical approximation would
only be able to approximate these contributions up to one loop. An
explicit calculation should decide on whether these contributions
occur or not. The sample calculation in \cite{Weinberg:2005vy} seems
to indicate that there are no late time contributions, even for large
internal momenta, but we are concerned about the fact that in this
calculation, $\tau \to 0$ is taken before the momentum integrals are
performed. The terms that are discarded in this way, might lead to
late time contributions.

Note that the background spacetime in inflation is not exactly de
Sitter, but typically has a slowly decreasing Hubble constant
$H$. This time dependence should be taken into account when deriving
the wavefunctions for the fluctuating fields, and when doing the time
integrals. Moreover, the fluctuations can react back on this
background and in this way change the time dependence of $H$. This
backreaction can be calculated by considering one point functions of
the fluctuating fields, as is done for example in \cite{Abramo:1998hj,
  Sloth:2006az, Sloth:2006nu}.

\subsection{Conclusions}
\label{conclusions}
We have investigated up to which order corrections to cosmological
correlation functions, generated after horizon exit, can be calculated
reliably using classical physics. We have done this by making a
detailed study of $\phi^3$ theory on a de Sitter background, for a
massless minimally coupled scalar field $\phi$, as a toy model.

In the quantum theory we studied late time contributions (generated
after horizon exit) to equal time correlation functions with external
momenta much smaller than the Hubble scale. We found that in loop
corrections, the loop integrals get contributions from internal
momenta up to the Hubble scale $H$. This is different from the
intuition from effective field theories in flat space, that loop
integrals are dominated by internal momenta of the same order of
magnitude as the external momenta. Our calculation of the one loop
correction to the two point function supports the argument that the
contributions from large internal momenta (around scale $H$) are not
negligible: they are proportional to $\lambda^2 \ln^3(\tau/\tau_H)$,
whereas the contributions from the small internal momenta are also
proportional to $\lambda^2 \ln^3(\tau/\tau_H)$, or to $\lambda^2 \ln^2
(\tau/\tau_H)/\epsilon$, with $\epsilon = m^2/3 H^2$. 

Furthermore we found that a classical approximation can approximate
contributions from small internal momenta quite well, but that this
does not hold for contributions from large internal momenta. This is
not surprising, because the classical approximation is only supposed
to work for physics at momentum scales much below the Hubble scale
$H$. As a consequence, the classical approximation is good at tree
level, but in general not for loop corrections. An exception is the
one loop correction, for which the classical approximation can be good
if an ultraviolet cutoff $\Lambda > H$ is introduced.

We argued that the results presented for the $\phi^3$ toy model can be
extended to derivative interactions, and be applied to the curvature
perturbation $\zeta$. For non-Gaussian effects in multifield inflation
models this means that at leading order, which is tree level, can be
approximated quite well using classical physics. Also the one loop
corrections can be approximated if a suitable cutoff is chosen. But
for higher order corrections, the classical approximation is not
expected to be good.

For small internal momenta, we found that derivatives tend to suppress
late time corrections, in a way that agrees with the theorem derived
by Weinberg \cite{Weinberg:2005vy}. However, for large internal
momenta, late time contributions need not to be suppressed.
Consequently, it is possible that the curvature perturbation $\zeta$
is not conserved to all orders after horizon exit, even for single
field inflation. There might be contributions to correlation functions
of $\zeta$ that grow after horizon exit, coming from loop
corrections. These contributions would be suppressed by powers of the
coupling constant $H/M_{\rm pl}$ and possibly also by slow roll
parameters, but they would be amplified by powers of the number of
e-folds $\ln a$.

\acknowledgments
We thank Leo Kampmeijer, Jan Pieter van der Schaar and Koenraad Schalm
for many useful discussions in a preliminary stage of this
project. This work is supported by FOM/NWO.

\section*{Note added}
Simultaneously with this work, \cite{Prokopec:2007ak} appeared on the
arXiv, in which the stochastic approach is extended to scalar quantum
electrodynamics. Similar as in \cite{Miao:2006pn} where the fermion is
integrated out, here the vector field is integrated out and the
resulting effective theory for the scalar is used for the stochastic
approach.

\appendix
\section{Free scalar field on a de Sitter background}
\label{appwavefunction}
\subsection{Scalar wavefunction}
The free field equation of motion for a scalar field on a de Sitter
background is
\begin{equation}
\partial_{\tau}^2 \phi(x)+ 2 H a(\tau) \partial_{\tau} \phi(x) -
\nabla^2 \phi(x) + a^2 (\tau) \Big(m^2 + \xi R \Big) \phi(x)= 0,
\end{equation}
where we use $x=(\tau,\mathbf{x})$ with $\tau$ conformal time. After
a spatial Fourier transformation the solutions for the mode functions
are (see e.g. \cite{Birrell:1982ix})
\begin{equation}
\phi_{k,\alpha}(\tau) = - \frac{\sqrt{-\pi \tau}}{2 a(\tau)}
H_{\nu}^{(\alpha)} (-k \tau), \label{hankel}
\end{equation}
where $H_{\nu}^{(\alpha)}(-k\tau)$ are the Hankel functions for
$\alpha=1,2$, and where $\nu$ is determined by
\begin{equation}
\nu^2 = \frac{9}{4}- \frac{m^2}{H^2} - 12 \xi. \label{nu}
\end{equation}
For massless minimally coupled fields, $m=0$, $\xi=0$ and we take $\nu
= 3/2$ (we choose $\nu$ to be positive). Then the modefunctions reduce
to
\begin{equation}
\phi_{k,1}(\tau) = i \frac{H}{\sqrt{2 k^3}} (1+i k \tau) e^{-ik
  \tau} \label{masslessmodefunctions} 
\end{equation}
and $\phi_{k,2}(\tau)= \phi_{k,1}^*(\tau)$. For $k\tau \to -\infty$
the $\phi_{k,1} (\tau)$ mode function is proportional to $e^{-i k
\tau}/a(\tau)$ and is called the positive frequency
solution\footnote{Often the Hankel functions are used with a negative
argument: $H^{(\alpha)}_{\nu}(k\tau)$. Then the $\phi_{k,2}(\tau)$ mode
function is the positive frequency solution.}. Using these mode
functions, the scalar field operator can be decomposed as
\begin{equation}
\phi(\tau,\mathbf{x}) = \int \frac{d^3 k}{(2 \pi)^3} \left( e^{i
  \mathbf{k} \cdot \mathbf{x}} \alpha_{\mathbf{k}} \,\phi_{k,1} (\tau)
  + e^{- i \mathbf{k} \cdot \mathbf{x}} \alpha_{\mathbf{k}}^{\dagger}
  \, \phi_{k,1}^* (\tau) \right), \label{modedecom}
\end{equation}
where the annihilation operators $\alpha_{\mathbf{k}}$ and creation
operators $\alpha_{\mathbf{k}}^{\dagger}$ satisfy the commutation relations
\begin{equation}
[\alpha_{\mathbf{k}}, \alpha^{\dagger}_{\mathbf{k}'}] = (2\pi)^3 \delta^3
(\mathbf{k} - \mathbf{k}'), \hspace{1cm} [\alpha_{\mathbf{k}},
  \alpha_{\mathbf{k}'}] = 0.
\end{equation}
The normalizations are chosen such that 
\begin{equation}
[\phi(\tau,\mathbf{x}),\pi(\tau,\mathbf{x}')] = i \delta^3(\mathbf{x}
- \mathbf{x}'),
\end{equation}
where $\pi(\tau,\mathbf{x})= a^2(\tau) \partial_{\tau}
\phi(\tau,\mathbf{x})$ is the conjugate momentum, and that $a(\tau)
\phi(\tau,\mathbf{x})$ is a conventionally normalized free field for
$k\tau \to -\infty$.  The state $|0\rangle$ defined by
\begin{equation}
\alpha_{\mathbf{k}} |0\rangle = 0
\end{equation}
corresponds therefore with the free vacuum state in Minkowski
spacetime for $k \tau \to -\infty$. This state is called adiabatic or
Bunch-Davies vacuum and is the state with respect to which we
calculate correlation functions in this paper.

\subsection{Particle creation}
Despite the confusing nature of the concept of particle number in
curved spacetime (see e.g. \cite{Birrell:1982ix}), we give here a
definition of the particle number in the frame of a comoving observer
in de Sitter spacetime. This definition of particle number then allows
us to make a comparison with the classical approximation in thermal
field theory in the next subsection.

Using comoving time, the free Lagrangian is the spatial integral over
the Lagrangian density \eqref{lagrangian} with $\lambda=0$
\begin{equation}
L[\varphi_{\mathbf{k}}, \partial_{\tau} \varphi_{\mathbf{k}}] = \int
\frac{d^3 k}{(2 \pi)^3} 
\left(\frac{1}{2} a^2 |\partial_{\tau} \varphi_{\mathbf{k}} |^2 -
\frac{1}{2} a^2 k^2 |\varphi_{\mathbf{k}}|^2 - \frac{1}{2} (m^2 + \xi
R) a^4 |\varphi_{\mathbf{k}}|^2 \right),
\end{equation}
where we have taken
\begin{equation}
\phi(\tau,\mathbf{x}) = \int \frac{d^3 k}{(2 \pi)^3} \,
  \varphi_{\mathbf{k}} \, e^{i \mathbf{k} \cdot \mathbf{x}},
\end{equation}
where the $\varphi_{\mathbf{k}}$ are time dependent operators.  In
this form the Lagrangian describes a system of uncoupled harmonic
oscillators with frequencies $\omega_k = \sqrt{k^2 + a^2( m^2 + \xi
R)}$. The conjugate momentum is defined as $\pi_{\mathbf{k}} =\delta
L/\delta (\partial_{\tau} \varphi_{\mathbf{k}}) = a^2\partial_{\tau}
\varphi_{\mathbf{k}} $, and using this the Hamiltonian becomes
\begin{equation}
H[\varphi_{\mathbf{k}},\pi_{\mathbf{k}}] = \int \frac{d^3 k}{(2
  \pi)^3} \left(\frac{1}{2} 
\frac{|\pi_{\mathbf{k}}|^2}{a^2} + \frac{1}{2} 
a^2 k^2 | \varphi_{\mathbf{k}} |^2 + \frac{1}{2} (m^2 + \xi R) a^4
|\varphi_{\mathbf{k}}|^2 \right). \label{dsham}
\end{equation}
By analogy to the harmonic oscillator, we define time dependent
annihilation and creation operators $\bar{\alpha}_{\mathbf{k}}$ and
$\bar{\alpha}_{\mathbf{k}}^{\dagger}$ by
\begin{equation}
a \varphi_{\mathbf{k}} = \frac{1}{\sqrt{2 \omega_k}} \Big(
    \bar{\alpha}_{\mathbf{k}} \, e^{-i k \tau} + 
    \bar{\alpha}_{-\mathbf{k}}^{\dagger} \, e^{i k \tau}\Big),
    \hspace{.05\textwidth} 
    \frac{\pi_{\mathbf{k}}}{a} = \frac{1}{i}  
    \sqrt{\frac{\omega_k}{2}} 
    \Big(\bar{\alpha}_{\mathbf{k}} \, e^{-i k \tau} -
    \bar{\alpha}_{-\mathbf{k}}^{\dagger} \, e^{i k \tau} \Big).
\end{equation}
They obey the
commutation relation
\begin{equation}
[\bar{\alpha}_{\mathbf{k}},\bar{\alpha}^{\dagger}_{\mathbf{k'}}]=
(2\pi)^3 \delta^3(\mathbf{k}- \mathbf{k}'). 
\end{equation}
The Hamiltonian \eqref{dsham} expressed in terms of these operators
has the familiar form
\begin{equation}
H = \int \frac{d^3 k}{(2 \pi)^3}
\Big(\bar{\alpha}_{\mathbf{k}}^{\dagger} \bar{\alpha}_{\mathbf{k}} +
\frac{1}{2}
 [\bar{\alpha}_{\mathbf{k}},\bar{\alpha}^{\dagger}_{\mathbf{k}} ] 
  \Big) \omega_k .
\end{equation}

If we take $m=0$ and $\xi=0$ we can use the modefunction $\phi_{k,1}
(\tau)$ of equation \eqref{masslessmodefunctions} to express the
operators $\bar{\alpha}_{\mathbf{k}}$,
$\bar{\alpha}_{\mathbf{k}}^{\dagger}$ in terms of the operators
$\alpha_{\mathbf{k}}$, $\alpha_{\mathbf{k}}^{\dagger}$ in the mode
decomposition \eqref{modedecom}:
\begin{align}
\bar{\alpha}_{\mathbf{k}} &= \frac{-i}{2 k \tau}(1+2 i k \tau)
\alpha_{\mathbf{k}} + i \frac{e^{2 i k \tau}}{2 k \tau}
\alpha_{-\mathbf{k}}^{\dagger}, \\
\bar{\alpha}_{-\mathbf{k}}^{\dagger} & = \frac{- i e^{-2 i k \tau}}{2 k
  \tau} \alpha_{\mathbf{k}} + \frac{i}{2 k \tau} (1 - 2i k
\tau) \alpha_{-\mathbf{k}}^{\dagger}.
\end{align}
For $k \tau  \to -\infty$ the $\bar{\alpha}_{\mathbf{k}}$ operator
becomes equal to $\alpha_{\mathbf{k}}$. We can define $n_k$ and
$\tilde{n}_k$ by
\begin{align}
\langle 0 | \bar{\alpha}_{\mathbf{k}}^{\dagger}
\bar{\alpha}_{\mathbf{k}'} |0 \rangle &= n_k \;
(2\pi)^3 \delta(\mathbf{k}-\mathbf{k}'), & n_k = \frac{1}{4 k^2
  \tau^2} \\
\langle 0 | \bar{\alpha}_{\mathbf{k}}
\bar{\alpha}_{-\mathbf{k}'} | 0 \rangle &= (\tilde{n}_k + \frac{i}{2
  k\tau}) (2 \pi)^3 \delta(\mathbf{k} - \mathbf{k}'), & \tilde{n}_k =
\frac{1}{4 k^2 \tau^2}
\end{align}
where one can interpret $n_k$ as the particle number and $\tilde{n}_k$
as a kind of off-diagonal particle number, with respect to the vacuum
at $k\tau \to - \infty$. Clearly these quantities are proportional to
$a^2$.

\subsection{Comparison with thermal field theory}
It is interesting to compare with thermal field theory on a Minkowski
background (see \cite{Aarts:1996qi,Aarts:1997kp}). The $F$ two point
function in a thermal system with temperature $T$ in Minkowski
spacetime is given by
\begin{equation}
F(k,t_1,t_2) = \frac{1}{k} \left(n_{\rm th}(k) + \frac{1}{2}
\right) \cos k (t_1 -t_2)
\hspace{.05\textwidth} n_{\rm th}(k) = \frac{1}{e^{k/T} -1},
\end{equation}
where $n_{\rm th}(k)$ is the particle number. For small momenta with
respect to the temperature $k \ll T$, $n_{\rm th} \simeq T/k$, which
becomes large and classical behaviour emerges. Moreover the $G^R$ two
point function does not have this amplification factor; it is given by
\begin{equation}
G^R(k,t_1,t_2) = \frac{\theta(t_1-t_2)}{k} \sin k (t_1-t_2) .
\end{equation}
Therefore a diagram containing a vertex with more than one dashed line
is suppressed with respect to the same diagram with a vertex with one
dashed line. 

To compare this with the de Sitter case, it is natural to consider $a
\phi_k(\tau)$. Then we have
\begin{equation}
a(\tau_1) a(\tau_2) F(k,\tau_1,\tau_2) = \frac{1}{k} \left[ \left(
\frac{1}{2 k^2 \tau_1 \tau_2} + \frac{1}{2} \right) \cos k (\tau_1 -
\tau_2) + \frac{\tau_1 - \tau_2}{2 k\tau_1 \tau_2} \sin k (\tau_1 -
\tau_2) \right].
\end{equation}
We see that for equal times $\tau_1=\tau_2=\tau$ this quantity grows
as
\begin{equation}
\frac{1}{2 k^2 \tau^2} + \frac{1}{2} \propto n+\tilde{n} +
\frac{1}{2}. 
\end{equation}
The quantity $a(\tau_1) a(\tau_2) G^R(k,\tau_1, \tau_2)$ does not have
this amplification factor for late times, as can be seen from
expansion for small $k\tau$. Therefore a diagram containing a vertex
with more than one dashed line is suppressed with respect to the same
diagram with a vertex with one dashed line, as in thermal field
theory.

Both in the de Sitter case as in thermal field theory, the arguments
given above explain why the classical approximation is good for small
physical internal momenta, i.e. $k/a \ll H$ ($|k\tau| \ll 1$) or $k
\ll T$. But, as we argue in this paper for the de Sitter case, for
large internal momenta ($\approx H$ or $\approx T$) problems arise for
the classical approximation, e.g. in the thermal case involving Hard
Thermal Loops \cite{Aarts:1999wj}.

\section{Amputated diagrams with no external dashed lines}
\label{appnoextgr}
The contribution of an amputated diagram with no external dashed
lines, as for example occurs in diagram \eqref{vanishingdiagram},
vanishes. The reason is that in such a diagram there is always a
closed loop of $G^R$ two point functions. This can be seen by picking
an arbitrary vertex, and from there following one of the dashed lines,
along the $G^R$ two point functions, from this vertex to a next
vertex. From this new vertex one can repeat this to go to the next
vertex. Because there is no external dashed line, this can be repeated
indefinitely while staying inside the diagram. Since there is only a
finite number of vertices in the diagram, one eventually ends up at a
vertex for the second time. Therefore there must be a closed loop of
$G^R$ two point functions in a diagram with no external dashed lines.

Because of the $\theta$-functions of the $G^R$ two point functions,
the times of the vertices of a closed loop of $G^R$ two point
functions have to be ordered. But in a closed loop of there is always
at least one $G^R$ two point function for which the $\theta$-function
vanishes, and therefore the complete diagram vanishes. Even if all the
internal times of the loop are equal, the diagram vanishes because the
$G^R$ two point function vanishes if the time arguments are equal.

\section{Correspondence between diagrams in quantum and
  classical theory}
\label{appdiagqcl}
In this Appendix we show that, if one chooses initial conditions such
that 
\begin{equation}
F_{\rm cl}(\mathbf{k},\tau_1,\tau_2) = F(\mathbf{k},\tau_1,\tau_2),
\label{fcleqfq} 
\end{equation}
the classical approximation reproduces the diagrams of the quantum
theory with only vertices with one dashed line.

Discarding the vertex with three dashed lines in the quantum
theory corresponds to discarding the term with $(\phi^{(2)})^3$ in
equation \eqref{lagph12}. An arbitrary equal time correlation
function, calculated up to order $n$ in the coupling $\lambda$, can
then be written as
\begin{equation}
\langle \phi(\tau,\mathbf{x}_1) \ldots \phi(\tau,\mathbf{x}_r)
\rangle = \langle \phi^{(1)} (\tau,\mathbf{x}_1) \ldots
\phi^{(1)}(\tau,\mathbf{x}_r) \frac{1}{n!} \left( \bar{S}_{\rm int}
\right)^n \rangle^{(0)}, \label{quantexpval}
\end{equation}
where the superscript $(0)$ denotes the free correlation function, and
where the modified interaction part of the action is given by
\begin{equation}
\bar{S}_{\rm int} = -\int_{-\infty}^{\tau} d\tau a^4(\tau) \int d^3 x
\, \frac{\lambda}{2!}\, (\phi^{(1)})^2 \phi^{(2)} . 
\end{equation}
On the right hand side of equation \eqref{quantexpval} the fields
$\phi^{(1)}$ and $\phi^{(2)}$ have to be contracted in all possible
ways: every $\phi^{(2)}$ is contracted with a $\phi^{(1)}$ to a
retarded propagator $G^R$, and the remaining $\phi^{(1)}$'s are
contracted with each other to $F$ two point functions. Suppose that we
do have contracted the $\phi^{(2)}$'s, but not yet the remaining
$\phi^{(1)}$'s. The correlation function can then be drawn as a number
of disconnected diagrams, in which the vertices are connected by $G^R$
two point functions and which have loose ends representing
$\phi^{(1)}$'s that are yet to be connected. In total there are $n$
vertices. Each disconnected diagram can be a tree diagram or a loop
diagram. A loop diagram with only $G^R$ two point functions vanishes,
as explained in Appendix \ref{appnoextgr}. Hence if one of the
disconnected diagrams contains a loop, the complete contraction does
not contribute to the correlation function, and for all non-vanishing
contractions the disconnected diagrams are tree diagrams. These tree
diagrams are the same tree diagrams in the classical theory that
represent the perturbative solutions $\phi_i$ in the classical theory
\eqref{pertsol}, where all the $\phi^{(1)}$'s in the quantum theory
correspond with free field solutions $\phi_0$ in the classical
theory. The symmetry factors are also equal because they arise in the
same way (the factor $1/n!$ is canceled by a factor $n!$ from the $n$
equivalent vertices). The remaining contractions of the $\phi^{(1)}$'s
in the quantum theory are equal to the contractions of the $\phi_0$'s
in the classical theory, because of equation \eqref{fcleqfq}.

Therefore the classical theory with the same couplings as the quantum
theory, and using initial conditions such that \eqref{fcleqfq} holds,
gives all the diagrams using only the vertex with one dashed line in
the quantum theory, up to vanishing diagrams. Hence this classical
theory reproduces the late time contributions for small internal
momenta.

\section{One loop correction to two point function}
\label{app1loopcorr}
In this appendix we calculate the one loop correction to the two point
function as given in equation \eqref{1loopdiagrams}. We first consider
the diagrams with one external dashed line (diagrams A and D), and
then the ones with two external dashed lines (diagrams B and C). The
complete result is given in equation \eqref{endresult}.

\subsection{Diagrams A and D}
We start with the diagrams with one external $G^R$ two point
function. First we calculate the amputated diagrams, and then attach
the external lines. The amputated diagrams are:
\begin{equation}
\picc{
\DashLine(5,20)(22,20){2}
\CArc(32,20)(10,90,180)
\DashCArc(32,20)(10,0,90){2}
\CArc(32,20)(10,180,360)
\Line(42,20)(60,20)
\Text(66,36)[rt]{A}
\LongArrowArcn(32,20)(14,140,40)
\Text(32,35)[cb]{$\mathbf{p}'$}
\LongArrowArc(32,20)(14,220,320)
\Text(32,5)[ct]{$\mathbf{p}$}
\LongArrow(6,26)(18,26)
\Text(12,30)[cb]{$\mathbf{k}$}
\Text(2,16)[lt]{$\tau_1$}
\Text(60,16)[rt]{$\tau_2$}
} \hspace{.1\textwidth}
\picc{
\DashLine(5,20)(22,20){2}
\CArc(25,20)(3,0,360)
\Line(22.9,22.1)(27.1,17.9)
\Line(22.9,17.9)(27.1,22.1)
\Line(28,20)(46,20)
\LongArrow(6,26)(18,26)
\Text(12,30)[cb]{$\mathbf{k}$}
\Text(56,36)[rt]{D}
\Text(2,16)[lt]{$\tau_1$}
\Text(50,16)[rt]{$\tau_2$}
}.
\end{equation}
The amputated version of diagram A is given by 
\begin{multline}
{\rm A}_{\rm amp}(k,\tau_1,\tau_2) = \frac{-i (-i \lambda)^2}{H^8
  \tau_1^4 \tau_2^4} \int \frac{d^3 p d^3 
  p'}{(2 \pi)^3} \, \delta^3(\mathbf{k} -\mathbf{p} - \mathbf{p}')
  G^R(p',\tau_1, \tau_2) F( 
  p,\tau_1, \tau_2) = \\
\frac{i \lambda^2}{(2 \pi)^2 k H^8 \tau_1^4 \tau_2^4}
  \int_0^{\infty} dp \, p \int_{|p-k|}^{p+k} dp' \, p' \,
  G^R(p',\tau_1, \tau_2) F(p,\tau_1, \tau_2)
,  \label{genexprA}
\end{multline}
where we have used the identity
\begin{equation}
\int d^3 p \,d^3 p'\, \delta^3(\mathbf{k}+\mathbf{p}+\mathbf{p}')
f(k,p,p') = \frac{2 \pi}{k} \int_0^{\infty} dp \, p \int_{|p-k|}^{p+k}
dp' \, p' f(k,p,p'). \label{intidentity}
\end{equation}
We will evaluate this integral below. For diagram D we see from
equations \eqref{ctfeyn} and \eqref{deltas} that it is equal to 
\begin{equation}
{\rm D}_{\rm amp} (k,\tau_1, \tau_2) = -i \, a^4(\tau_1) \delta_m
 \delta(\tau_1 - \tau_2) = \frac{-i 
 \lambda^2}{4 (2 \pi)^2 H^4 \tau_1^4} \ln \left(\frac{\Lambda}{\mu}
 \right) \, \delta(
 \tau_1 - \tau_2), \label{contrct}
\end{equation}
where $\Lambda$ is the ultraviolet momentum cutoff and $\mu$ is a
renormalization scale. The counterterm $\delta_Z$ is finite and leads
to terms proportional to positive powers of $\tau$, and is therefore
left out. 

We calculate the integral \eqref{genexprA} by splitting the $p$
integral in a small momentum part $\int_0^{M_{\rm cm}} dp$ and a large
momentum part $\int_{M_{\rm cm}}^{\Lambda a(\tau_2)} dp$, with
$|M_{\rm cm} \tau_i| \ll 1$ and $M_{\rm cm} > k$.\footnote{In
\cite{Boyanovsky:2004ph} a similar split of integrals is used to
calculate a similar integral. Note however that the integral there
differs from the integral here, because the self-energy kernel of
\cite{Boyanovsky:2004ph} is not the same as the amputated diagram A.}

\subsubsection{Amputated diagram for small internal momenta}
\label{smampad}
The integral in equation \eqref{genexprA} is infrared divergent for
$p\to 0$. We regulate this divergence by giving the field a small mass
$m \ll H$, such that $\nu=3/2-\epsilon$ with $\epsilon =
m^2/3H^2$. The $F$ and $G^R$ two point functions are then, using
equations \eqref{qfp}, \eqref{qrp} and \eqref{modedecom},
\begin{align}
F(k,\tau_1,\tau_2) &= \frac{\pi \sqrt{\tau_1 \tau_2}}{4 a(\tau_1)
  a(\tau_2)} \, {\rm Re} \Big(H^{(1)}_{\nu} (-k\tau_1)
  {H^{(1)}_{\nu}}^* (-k\tau_2) \Big), \\ 
G^R(k,\tau_1,\tau_2) &=
  -\frac{\pi \sqrt{\tau_1 \tau_2}}{2 a(\tau_1) a(\tau_2)}
  \theta(\tau_1 - \tau_2) \, {\rm Im} \Big(H^{(1)}_{\nu} (-k\tau_1)
  {H^{(1)}_{\nu}}^* (-k\tau_2) \Big).
\end{align}
Using (see \cite{Abramowitz:1965aa})
\begin{align}
H^{(1)}_{\nu} (-k\tau) &= J_{\nu}(-k\tau) + i \left( \frac{\cos \nu
  \pi}{\sin \nu \pi} J_{\nu}(- k \tau) - \frac{1}{\sin \nu \pi}
  J_{-\nu} (-k\tau) \right), \\
J_{\nu}(-k\tau) &= \frac{1}{\Gamma(\nu+1)} \big(-\frac{1}{2} k \tau
  \big)^{\nu} \Big(1 + \mathcal{O}(k^2 \tau^2) \Big),
\end{align}
and the identity $\Gamma(\nu) \Gamma(1-\nu) = \pi/\sin \nu \pi$, we
obtain
\begin{align}
F(k,\tau_1,\tau_2) & = \frac{H^2}{2 k^3}(k^2 \tau_1
\tau_2)^{\epsilon}, \label{expandedf} \\
G^R(k,\tau_1,\tau_2) &= \theta(\tau_1 - \tau_2)  \frac{H^2}{3}
\left(\tau_1^{3} \left(\frac{\tau_2}{\tau_1}\right)^{\epsilon} -
\left(\frac{\tau_1}{\tau_2}\right)^{\epsilon} \tau_2^{3}
\right). \label{expandedgr}  
\end{align}

The integral is 
\begin{align}
\frac{i \lambda^2}{(2 \pi)^2 k H^8 \tau_1^4 \tau_2^4}
\frac{H^4 \theta(\tau_1-\tau_2) \left(\tau_1^3 |\tau_2|^{2 \epsilon} -
  |\tau_1|^{2 \epsilon} \tau_2^3 \right)}{6} \int_0^{M_{\rm cm}} dp \, \frac{p^{2
    \epsilon}}{p^2} \int_{|p-k|}^{p+k} dp' \, p' &= \nonumber \\
\frac{i \lambda^2 \theta(\tau_1-\tau_2) \left(\tau_1^3 |\tau_2|^{2 \epsilon} -
  |\tau_1|^{2 \epsilon} \tau_2^3 \right)}{6 (2\pi)^2 H^4 (\tau_1
  \tau_2)^4} \frac{{M_{\rm cm}}^{2 \epsilon}}{\epsilon} &= \nonumber \\ \frac{i \lambda^2
  \theta(\tau_1-\tau_2)}{6 (2\pi)^2 H^4 (\tau_1 \tau_2)^4} \left(
\frac{\tau_1^3 - \tau_2^3}{ \epsilon} + 2 \tau_1^3 \ln |{M_{\rm cm}}
\tau_2| - 2 \tau_2^3 \ln |{M_{\rm cm}} \tau_1| + \mathcal{O}(\epsilon)
\right).& \label{contrsm}
\end{align}

\subsubsection{Amputated diagrams for large internal momenta}
\label{largemomentadiaga}
For large momenta we approximate the field to be massless and we use
the two point functions of equations \eqref{freef} and
\eqref{freegr}, which we write as
\begin{align}
F(k,\tau_1,\tau_2) &= \frac{H^2}{2} \sum_{i=1}^3 F_i (k,\tau_1,\tau_2),
\label{firstlinef} \\ 
F_1(k,\tau_1,\tau_2) & = \frac{1}{k^3} \cos k \Delta \tau, \nonumber \\
F_2(k,\tau_1,\tau_2) &= \frac{1}{k^2} \Delta \tau \sin k \Delta
\tau, \nonumber \\
F_3(k,\tau_1,\tau_2) &= \frac{1}{k} \tau_1 \tau_2 \cos k \Delta
\tau , \nonumber
\end{align}
with $\Delta \tau = \tau_1-\tau_2$, and similarly
\begin{align}
G^R(k,\tau_1,\tau_2) &= \theta(\tau_1-\tau_2) H^2 \sum_{i=1}^3 G^R_i
(k,\tau_1,\tau_2), \\ 
G^R_1(k,\tau_1,\tau_2) & = \frac{1}{k^3} \sin k \Delta \tau, \nonumber
\\
G^R_2(k,\tau_1,\tau_2) &= \frac{-1}{k^2} \Delta \tau \cos k \Delta
\tau, \nonumber  \\
G^R_3(k,\tau_1,\tau_2) &= \frac{1}{k} \tau_1 \tau_2 \sin k \Delta
\tau . \label{lastlinegr} \nonumber
\end{align}
In the following calculations we use the definitions
\begin{equation}
{\rm Si}(x) = \int_0^x dx' \frac{\sin x'}{x'}, \hspace{.1\textwidth}
{\rm Ci}(x) = - \int_x^{\infty} dx' \frac{\cos x'}{x'},
\end{equation}
which behave for small respectively large arguments as
\begin{align}
& {\rm Si}(x) = x + \mathcal{O}(x^3), & {\rm
    Si}(x) = \frac{\pi}{2}- \frac{\cos x}{x} -\frac{\sin x}{x^2} +
    \mathcal{O}(x^{-3}), \\  
& {\rm Ci}(x) = \gamma + \ln x -\frac{x^2}{4} + \mathcal{O}(x^4),
   &  {\rm Ci}(x) = \frac{\sin x}{x} -\frac{\cos x}{x^2} +
    \mathcal{O} (x^{-2}),
\end{align}
and the identities
\begin{align}
\int_{p-k}^{p+k} dp' \, \frac{\sin p' \Delta \tau}{p'^2}  = & -
\frac{\sin (p+k) \Delta \tau}{p+k} + \frac{\sin (p-k)\Delta \tau}{p-k}
+ \nonumber \\ & \hspace{.1\textwidth} \Delta \tau \Big({\rm
  Ci}((p+k)\Delta \tau) - 
 {\rm Ci} ((p-k) \Delta \tau) \Big), \\
\int_{p-k}^{p+k} dp' \, \frac{\cos p' \Delta \tau}{p'^2} = & -
\frac{\cos (p+k) \Delta \tau}{p+k} + \frac{\cos(p-k) \Delta \tau}{p-k}
+ \nonumber \\
& \hspace{.1\textwidth} - \Delta \tau \Big( {\rm Si}((p+k)\Delta
\tau) - {\rm Si}((p-k) \Delta \tau) \Big), \\
\int_{p-k}^{p+k} dp'\, \sin p' \Delta \tau = & \frac{-1}{\Delta \tau}
\left( \cos(p+k) \Delta \tau - \cos (p-k) \Delta \tau \right) =
\frac{2}{\Delta \tau} \sin k \Delta \tau \sin p \Delta\tau, \\
\int_{p-k}^{p+k} dp'\, \cos p' \Delta \tau = & \frac{1}{\Delta \tau}
\left( \sin(p+k) \Delta \tau - \sin (p-k) \Delta \tau \right) =
\frac{2}{\Delta \tau} \sin k \Delta \tau \cos p \Delta\tau .
\end{align}

Next we calculate the contributions
\begin{equation}
\int_{M_{\rm cm}}^{\Lambda a(\tau_2)} dp\, p \, F_i(p,\tau_1, \tau_2)
\int_{p-k}^{p+k} dp' \, p' \,G^R_j(p',\tau_1, \tau_2), \label{ijcontrs}
\end{equation}
for $i$ and $j$ from $1$ to $3$:
\begin{itemize}
\item[\#1:] $F_1(p,\tau_1,\tau_2) G^R_1(p',\tau_1,\tau_2)$
\begin{align}
& \int_{M_{\rm cm}}^{\Lambda a(\tau_2)} dp\, \frac{\cos p \Delta\tau}{p^2}
\int_{p-k}^{p+k} dp'\, \frac{\sin p' \Delta \tau}{p'^2} = \nonumber\\
& \hspace{.1\textwidth} \bigg[-\frac{\cos p\Delta\tau}{p}
  \bigg(-\frac{\sin(p+k)\Delta \tau}{ 
    p+k} + \frac{\sin(p-k) \Delta \tau}{p-k} + \nonumber \\ &
  \hspace{.3\textwidth} \Delta \tau \Big( {\rm
    Ci}((p+k) \Delta \tau) - {\rm Ci}((p-k) \Delta \tau) \Big)
  \bigg) \bigg]^{\Lambda a(\tau_2)}_{M_{\rm cm}} + & \nonumber \\
& \hspace{.1\textwidth} \int_{M_{\rm cm}}^{\Lambda a(\tau_2)} dp \bigg\{ -\Delta
    \tau \frac{\sin p \Delta 
  \tau}{p} \bigg(- \frac{\sin(p+k) \Delta \tau}{p+k} + \frac{\sin
  (p-k) \Delta \tau}{p-k} + \nonumber \\ & \hspace{.3\textwidth} \Delta \tau
   \Big( {\rm Ci}((p+k) \Delta \tau) - 
 {\rm Ci}((p-k) \Delta \tau) \Big) \bigg) + \nonumber \\ &
    \hspace{.3\textwidth} \frac{\cos p \Delta
  \tau}{p} \bigg( \frac{\sin (p+k) \Delta \tau}{(p+k)^2} - \frac{ \sin
  (p-k) \Delta \tau}{(p-k)^2} \bigg) \bigg\}, & \label{lmcontr1}
\end{align}
\item[\#2:] $F_1(p,\tau_1,\tau_2) G^R_2(p',\tau_1,\tau_2)$
\begin{align}
& -\Delta \tau \int_{M_{\rm cm}}^{\Lambda a(\tau_2)} dp\, \frac{\cos p \Delta\tau}{p^2}
\int_{p-k}^{p+k} dp'\, \frac{\cos p' \Delta \tau}{p'} = \nonumber \\
& \hspace{.15\textwidth} \Delta \tau \bigg[ \frac{\cos p\Delta \tau}{p}
  \Big( {\rm Ci}((p+k) \Delta \tau) - {\rm Ci}((p-k)\Delta \tau)
  \Big) \bigg]^{\Lambda a(\tau_2)}_{M_{\rm cm}} + \nonumber \\
& \hspace{.15\textwidth} \Delta \tau \int_{M_{\rm cm}}^{\Lambda a(\tau_2)} dp\,
\bigg\{ \Delta \tau 
\frac{\sin p\Delta \tau}{p} \Big( {\rm Ci}((p+k)\Delta \tau) - {\rm
  Ci}(( p-k)\Delta\tau) \Big)+ \nonumber \\ & \hspace{.3\textwidth}
-\frac{\cos p\Delta\tau}{p} \bigg( 
\frac{\cos (p+k)\Delta \tau}{p+k} - \frac{\cos(p-k) \Delta \tau}{ p-k}
\bigg) \bigg\},
\end{align}
\item[\#3:] $F_1(p,\tau_1,\tau_2) G^R_3(p',\tau_1,\tau_2)$
\begin{align}
& \tau_1 \tau_2 \int_{M_{\rm cm}}^{\Lambda a(\tau_2)} dp\, \frac{\cos p \Delta\tau}{p^2}
\int_{p-k}^{p+k} dp'\, \sin p' \Delta \tau = & \nonumber \\
& \hspace{.1\textwidth} \frac{\tau_1 \tau_2 \sin k \Delta \tau}{\Delta
  \tau} \int_{M_{\rm cm}}^{\Lambda 
  a(\tau_2)} dp \, \frac{\sin 2 p \Delta \tau}{p^2} = & \nonumber \\
& \hspace{.3\textwidth} \frac{\tau_1 \tau_2 \sin k \Delta \tau}{\Delta
  \tau} \bigg[ - \frac{ 
    \sin 2 p \Delta \tau}{p} + 2 \Delta \tau \, {\rm Ci}(2 p \Delta \tau)
  \bigg]^{\Lambda a(\tau_2)}_{M_{\rm cm}} ,
\end{align}
\item[\#4:] $F_2(p,\tau_1,\tau_2) G^R_1(p',\tau_1,\tau_2)$
\begin{align}
& \Delta \tau \int_{M_{\rm cm}}^{\Lambda a(\tau_2)} dp\, \frac{\sin p \Delta\tau}{p}
\int_{p-k}^{p+k} dp'\, \frac{\sin p' \Delta \tau}{p'^2} = \nonumber \\
& \hspace{.15\textwidth} \int_{M_{\rm cm}}^{\Lambda a(\tau_2)} dp \, \Delta
\tau \frac{\sin p \Delta 
  \tau}{p} \bigg(- \frac{\sin(p+k) \Delta \tau}{p+k} + \frac{\sin
  (p-k) \Delta \tau}{p-k} + \nonumber \\ & \hspace{.35\textwidth}
\Delta \tau \Big( 
     {\rm Ci}((p+k) \Delta \tau) - {\rm Ci}((p-k) \Delta \tau) \Big)
     \bigg), 
\end{align}
\item[\#5:] $F_2(p,\tau_1,\tau_2) G^R_2(p',\tau_1,\tau_2)$
\begin{multline}
-\Delta \tau^2 \int_{M_{\rm cm}}^{\Lambda a(\tau_2)} dp\, \frac{\sin p \Delta\tau}{p}
\int_{p-k}^{p+k} dp'\, \frac{\cos p' \Delta \tau}{p'} = \\
-\Delta \tau^2 \int_{M_{\rm cm}}^{\Lambda a(\tau_2)} dp\, \frac{\sin p\Delta
  \tau}{p} \Big( {\rm Ci}((p+k)\Delta \tau) - {\rm Ci}((p-k)\Delta
\tau) \Big), 
\end{multline}
\item[\#6:] $F_2(p,\tau_1,\tau_2) G^R_3(p',\tau_1,\tau_2)$
\begin{align}
& \Delta \tau \, \tau_1 \tau_2 \int_{M_{\rm cm}}^{\Lambda a(\tau_2)} dp\,
\frac{\sin p\Delta \tau}{p} \int_{p-k}^{p+k} dp' \, \sin p' \Delta
\tau = \nonumber \\ & \hspace{.1\textwidth} \tau_1 \tau_2 \sin k \Delta \tau
\int_{M_{\rm cm}}^{\Lambda a(\tau_2)} 
dp\, \frac{1- \cos 2 p \Delta \tau}{p} = \nonumber \\
& \hspace{.3\textwidth} \tau_1 \tau_2 \sin k\Delta \tau \bigg[ \ln p
- {\rm Ci}(2 p \Delta \tau) \bigg]^{\Lambda a(\tau_2)}_{M_{\rm cm}}, 
\end{align}
\item[\#7:] $F_3(p,\tau_1,\tau_2) G^R_1(p',\tau_1,\tau_2)$
\begin{align}
& \tau_1 \tau_2 \int_{M_{\rm cm}}^{\Lambda a(\tau_2)} dp\, \cos p \Delta\tau
\int_{p-k}^{p+k} dp'\, \frac{\sin p' \Delta \tau}{p'^2} = \nonumber \\
& \hspace{.05\textwidth} \tau_1 \tau_2 \int_{M_{\rm cm}}^{\Lambda a(\tau_2)} dp\,
\cos p \Delta \tau 
\bigg( -\frac{\sin (p+k) \Delta \tau}{p+k} + \frac{\sin (p-k) \Delta
  \tau}{p-k} + \nonumber \\ & \hspace{.4\textwidth} \Delta \tau \Big(
     {\rm Ci}((p+k) \Delta \tau) - 
     {\rm Ci}((p-k) \Delta \tau) \Big) \bigg) & = \nonumber \\
&\hspace{.05\textwidth} -\frac{\tau_1 \tau_2}{2} \bigg[ \sin k \Delta
       \tau \Big( \ln(p^2 - k^2) - {\rm Ci}(2(p+k)\Delta \tau) - {\rm
	 Ci}( 2(p-k) \Delta \tau) \Big) + \nonumber \\
&\hspace{.25\textwidth} \cos k \Delta \tau \Big( {\rm Si}(2(p +k)
       \Delta \tau) - {\rm Si}(2(p-k) \Delta \tau) \Big)
       \bigg]_{M_{\rm cm}}^{\Lambda a(\tau_2)} + \nonumber \\ 
& \hspace{.15\textwidth} \tau_1 \tau_2 \, \Delta \tau \int_{M_{\rm cm}}^{\Lambda
   a(\tau_2)} dp\, \cos p \Delta \tau \Big({\rm Ci}((p+k) \Delta \tau) - 
     {\rm Ci}((p-k) \Delta \tau) \Big),  
\end{align}
\item[\#8:] $F_3(p,\tau_1,\tau_2) G^R_2(p',\tau_1,\tau_2)$
\begin{multline}
-\tau_1 \tau_2  \Delta \tau \, \int_{M_{\rm cm}}^{\Lambda a(\tau_2)} dp\,
\cos p \Delta\tau \int_{p-k}^{p+k} dp'\, \frac{\cos p' \Delta
  \tau}{p'} = \\ 
-\tau_1 \tau_2 \Delta \tau \int_{M_{\rm cm}}^{\Lambda a(\tau_2)} dp \, \cos p
\Delta \tau \Big( {\rm Ci}((p+k) \Delta \tau) - {\rm Ci}((p-k)\Delta
\tau) \Big),
\end{multline}
\item[\#9:] $F_3(p,\tau_1,\tau_2) G^R_3(p',\tau_1,\tau_2)$
\begin{equation}
\tau_1^2 \tau_2^2 \int_{M_{\rm cm}}^{\Lambda a(\tau_2)} dp\,
\cos p\Delta \tau \int_{p-k}^{p+k} dp' \, \sin p' \Delta \tau = 
\frac{\tau_1^2 \tau_2^2}{\Delta \tau^2} \sin k\Delta \tau \bigg[
\sin^2 p \Delta \tau \bigg]^{\Lambda a(\tau_2)}_{M_{\rm cm}}. \label{lmcontr9}
\end{equation}
\end{itemize}
Together this becomes
\begin{align}
& \bigg[\frac{\cos p\Delta\tau}{p}
  \bigg(\frac{\sin(p+k)\Delta \tau}{ 
    p+k} - \frac{\sin(p-k) \Delta \tau}{p-k} \bigg) + \nonumber \\
& \hspace{.05\textwidth} \frac{\tau_1 \tau_2 \sin k \Delta \tau}{2}
  \bigg( 2 \,{\rm Ci}(2 p \Delta 
  \tau) + {\rm Ci}(2(p+k) \Delta \tau) + {\rm Ci}(2(p-k) \Delta \tau)
  + \nonumber \\ & \hspace{.3\textwidth} 
\ln \frac{p^2}{p^2 - k^2} - 2 \frac{ 
    \sin 2 p \Delta \tau}{p \Delta \tau} \bigg) + \nonumber \\ 
&\hspace{.05\textwidth} -\frac{\tau_1 \tau_2}{2} \cos k \Delta \tau
  \Big( {\rm Si}(2(p+k) \Delta \tau) - {\rm Si}(2(p-k) \Delta \tau)
  \Big) + 
\frac{\tau_1^2 \tau_2^2}{\Delta
  \tau^2} \sin k\Delta \tau \sin^2 p \Delta \tau 
\bigg]^{\Lambda a(\tau_2)}_{M_{\rm cm}} +
  \nonumber \\ 
& \int_{M_{\rm cm}}^{\Lambda a(\tau_2)} dp \; \frac{\cos p
  \Delta \tau}{p} 
  \bigg\{ \frac{\sin (p+k) \Delta \tau}{(p+k)^2} - \frac{ \sin 
  (p-k) \Delta \tau}{(p-k)^2} + \nonumber \\ & \hspace{.3\textwidth}
  -\Delta \tau \bigg( \frac{\cos 
  (p+k)\Delta \tau}{p+k} - \frac{\cos(p-k) \Delta \tau}{ p-k} \bigg)
  \bigg\} . \label{verylongeq}
\end{align}
For the upper limit the boundary term vanishes as $1/\Lambda^2$,
except the last term which we will discuss below. The lower limit of
the boundary term gives (where we use that $|{M_{\rm cm}} \tau_i |\ll 1$)
\begin{equation}
\frac{2}{3} k \Delta \tau^3 - 2 k \tau_1 \tau_2 \,\Delta \tau \Big(-2 +
\gamma + \ln 2 {M_{\rm cm}} \Delta \tau \Big) + \mathcal{O}(\tau_i^4).
\end{equation}
Using Mathematica, the integral in \eqref{verylongeq} becomes for $|{M_{\rm cm}}
\tau_i | \ll 1$
\begin{equation}
\frac{k \Delta \tau^3}{9} \Big( 8 - 6 \gamma - 6 \ln 2 {M_{\rm cm}} \Delta \tau
\Big) + \mathcal{O}(\tau_i^4).
\end{equation}
Together equation \eqref{verylongeq} becomes
\begin{equation}
\frac{2k}{3} (\tau_1^3 -\tau_2^3) \Big(\frac{7}{3}-\gamma - \ln 2 {M_{\rm cm}} (\tau_1
- \tau_2) \Big) -\frac{2}{3} k \tau_1 \tau_2 (\tau_1 - \tau_2)
+ \frac{\tau_1^2 \tau_2^2}{\Delta \tau^2} \sin k \Delta \tau \,
\sin^2 \Lambda a(\tau_2) \Delta \tau + \mathcal{O}(\tau_i^4)
. \label{contrlm}
\end{equation}

The term that contains the $\sin^2 \Lambda$ is logarithmically
divergent for $\Lambda \to \infty$. This can be seen as
follows. Consider the integral
\begin{equation}
\int_{-\infty}^{\infty} d\Delta\tau \, \theta(\Delta\tau) \, f(\Delta
  \tau) \frac{\sin^2 \Lambda a(\tau_2) \Delta \tau}{\Delta \tau} =
  \frac{1}{2} \int_{0}^{\infty} d\Delta\tau \, f(\Delta \tau)
  \frac{1-\cos \left(\frac{ -2\Lambda}{H} \frac{\Delta \tau}{\tau_1 -
  \Delta \tau} \right)}{ \Delta \tau},
\end{equation}
where $f(\Delta \tau)$ is a test function. The integral can be split
up into two integrals
\begin{equation}
\int_0^{\infty} = \lim_{\varepsilon\to 0} \int_{\varepsilon}^{\eta} +
\int_{\eta}^{\infty},
\end{equation}
where $\eta$ is used as a regulator time, which we take to zero in the
end, after taking the limit $\Lambda \to \infty$. In the first
integral we can approximate
\begin{equation}
\frac{\Delta \tau}{\tau_1 - \Delta \tau} \approx \frac{\Delta \tau}{
  \tau_1} ,\hspace{1cm} f(\Delta \tau) \approx f(0),
\end{equation}
so that it becomes
\begin{multline}
\lim_{\varepsilon\to 0} \int_{\varepsilon}^{\eta} d\Delta \tau \, f(0)
\frac{1-\cos \left(\frac{ -2\Lambda}{H} \frac{\Delta \tau}{\tau_1}
  \right)}{ \Delta \tau} = \lim_{\varepsilon\to 0}
f(0) \left(\ln\frac{\eta}{\varepsilon} -
    {\rm Ci}(\frac{-2 \Lambda \eta}{H \tau_1} ) + {\rm Ci} (\frac{ -2
      \Lambda \varepsilon}{H \tau_1}) \right) = \\
f(0) \left(\gamma + \ln \frac{-2 \Lambda \eta}{H\tau_1} \right),
\end{multline}
where we have taken ${\rm Ci}(-2 \Lambda \eta/H\tau_1) \to 0$, and
${\rm Ci}(-2 \Lambda \varepsilon/H\tau_1) \to \gamma + \ln(-2 \Lambda
\varepsilon / H\tau_1)$. 
The remaining integral is
\begin{equation}
\lim_{\Lambda \to \infty} \int_{\eta}^{\infty} d\Delta\tau\, f(\Delta
  \tau) \frac{1- \cos 
  (\frac{ -2\Lambda}{H} \frac{\Delta \tau}{\tau_1 - \Delta \tau})}{
  \Delta \tau} = \int_{\eta}^{\infty} d\Delta\tau\, \frac{f(\Delta
  \tau)}{ \Delta \tau},
\end{equation}
where the term with the cosine vanishes, provided that the test
function $f(\Delta \tau)$ vanishes sufficiently fast as $\Delta \tau
\to \infty$. Together we obtain for $\Lambda \to \infty$
\begin{multline}
\int_{-\infty}^{\infty} d\Delta \tau\, \theta(\Delta \tau) \, f(\Delta
  \tau) \frac{\sin^2 \Lambda 
  a(\tau_2) \Delta \tau}{\Delta \tau} = \\ \int_{-\infty}^{\infty}
  d\Delta \tau\, f(\Delta \tau)
  \frac{1}{2} \bigg[\frac{\theta(-\eta+\Delta\tau)}{\Delta \tau} +
  \delta(\Delta \tau) \left(\gamma + \ln\frac{-2 \Lambda \eta}{H\tau_1}
  \right) \bigg], \label{extrdiv} 
\end{multline}
which is in the language of distributions
\begin{equation}
 \theta(\Delta \tau) \frac{\sin^2 \Lambda a(\tau_2) \Delta
  \tau}{\Delta \tau} = \frac{1}{2} \bigg[\frac{ \theta(-\eta+\Delta
  \tau)}{ \Delta \tau} + \delta (\Delta \tau) \left(\gamma + \ln
  \frac{ -2 \Lambda \eta}{ H \tau_1} \right) \bigg].
\end{equation}

Using this result in equation \eqref{contrlm}, gathering the right
prefactors and adding the contribution from the counterterm
\eqref{contrct}, we obtain for the large momentum contribution
\begin{multline}
\frac{i \lambda^2 \theta(\tau_1 - \tau_2)}{2 (2 \pi)^2 H^4 (\tau_1
  \tau_2)^4} \bigg( 
\frac{2}{3} (\tau_1^3 - \tau_2^3)
  \Big(\frac{7}{3}-\gamma - \ln 2 {M_{\rm cm}} \Delta \tau \Big) - \frac{2}{3}
  \tau_1 \tau_2 (\tau_1 -\tau_2) + \\
\frac{(\tau_1 \tau_2)^2}{2} \bigg[\frac{\theta(-\eta + \Delta \tau)}{
  \Delta \tau} + \delta (\Delta \tau) \Big(\gamma + \ln \frac{-2 \mu
  \eta}{ H \tau_1} \Big) \bigg] \bigg). \label{resdiagalarge}
\end{multline}

\subsubsection{Attaching the external lines}
\label{extad}
Adding the small and large momenta contributions, we obtain for the
amputated diagrams A and D:
\begin{multline}
{\rm A}_{\rm amp}(k,\tau_1, \tau_2) + {\rm D}_{\rm amp}(k,
\tau_1, \tau_2) = 
\frac{i \lambda^2 \theta(\tau_1 - \tau_2)}{6 (2 \pi)^2 H^4 (\tau_1
  \tau_2)^4} \Bigg( (\tau_1^3-\tau_2^3) \bigg(\frac{1}{\epsilon} +
\frac{14}{3} - 2 \gamma \bigg) + \\ - 2 \tau_1 \tau_2 (\tau_1 - \tau_2)  +
2 \tau_1^3 \ln \left| \frac{\tau_2}{2(\tau_1 - \tau_2)} \right| - 2
\tau_2^3 \ln \left| \frac{\tau_1}{2(\tau_1 - \tau_2)} \right| + \\
\frac{3}{2} (\tau_1 \tau_2)^2 \bigg[ \frac{\theta(-\eta+\tau_1 -
    \tau_2)}{\tau_1 - \tau_2} + \delta (\tau_1 - \tau_2) \Big(\gamma +
  \ln \frac{-2 \mu \eta}{H \tau_1} \Big) \bigg] +
\mathcal{O}(\tau_i^4) + \mathcal{O}(\epsilon) \Bigg),
\end{multline}
where the dependence on ${M_{\rm cm}}$ has dropped out. The full correlation
function is obtained by
\begin{equation}
-i \int_{\tau_H}^{\tau} d\tau_1 \int_{\tau_H}^{\tau} d\tau_2 \,
G^R(k,\tau,\tau_1) F(k,\tau,\tau_2) \Big( {\rm A}_{\rm
  amp}(k,\tau_1,\tau_2) + {\rm D}_{\rm
  amp}(k,\tau_1,\tau_2) \Big).
\end{equation}
Because the external momentum $k$ is small, i.e. $|k\tau_i| \ll 1$, we
can use the expanded versions of the two point functions
\eqref{expanf}, \eqref{expangr} (or the ones of \eqref{expandedf},
\eqref{expandedgr}, but this gives only corrections of order
$\mathcal{O} (\epsilon)$). Using the integrals
\begin{align}
& \int_{\tau_H}^{\tau}d\tau_1 \int_{\tau_H}^{\tau_1} d\tau_2
\frac{(\tau^3- \tau_1^3) (\tau_1^3-\tau_2^3)}{(\tau_1 \tau_2)^4} =
\frac{1}{3} \left(1+ 2 \ln \frac{\tau}{\tau_H} + \frac{3}{2} \ln^2
\frac{\tau}{\tau_H} \right) +
\mathcal{O}(\frac{\tau}{\tau_H}), \\
& \int_{\tau_H}^{\tau}d\tau_1 \int_{\tau_H}^{\tau_1} d\tau_2
\frac{(\tau^3- \tau_1^3) (\tau_1-\tau_2)}{(\tau_1 \tau_2)^3} =
-\frac{1}{12} \left(11 + 6 \, \ln \frac{\tau}{\tau_H} \right) +
\mathcal{O} (\frac{\tau}{\tau_H}), \\
& \int_{\tau_H}^{\tau}d\tau_1 \int_{\tau_H}^{\tau_1} d\tau_2 \frac{\left(
\tau^3 -\tau_1^3 \right)}{(\tau_1 \tau_2)^4} \left(\tau_1^3 \ln \left|
\frac{\tau_2}{2(\tau_1 - \tau_2)} \right| - \tau_2^3 \ln \left|
\frac{\tau_1}{2(\tau_1 - \tau_2)} \right| \right)  = \frac{1}{18}
\bigg(\frac{97}{6}- 18\, \zeta(3) + \nonumber \\
& \hspace{.05\textwidth}  -2 \pi^2 -6 \ln 2 + (13-3 \pi^2 - 12 \ln 2) 
\ln \frac{\tau}{\tau_H} + (3-9 \ln 2) \ln^2 \frac{\tau}{\tau_H} + 3 \ln^3
\frac{\tau}{\tau_H} \bigg) + \mathcal{O}(\frac{\tau}{\tau_H}), \\
& \int_{\tau_H}^{\tau}d\tau_1 \int_{\tau_H}^{\tau_1} d\tau_2
\frac{(\tau^3-\tau_1^3)}{(\tau_1 \tau_2)^2} \bigg[ \frac{\theta(-\eta+\tau_1 -
    \tau_2)}{\tau_1 - \tau_2} + \delta (\tau_1 - \tau_2) \Big(\gamma +
  \ln \frac{-2 \mu \eta}{H \tau_1} \Big) \bigg] = \nonumber \\
& \hspace{.05\textwidth} \frac{1}{6} \bigg(8-2 \gamma-\pi^2 -2 \ln
\frac{2 \mu}{H} + 6\Big(1-\gamma -\ln \frac{2 \mu}{H}\Big) \ln
\frac{\tau}{\tau_H} \bigg) + \mathcal{O}(\frac{\tau}{\tau_H}),
\end{align}
(recall that $\eta$ is sent to zero), this becomes
\begin{align}
\frac{\lambda^2}{36(2 \pi)^2 k^3} \bigg\{ 
&\frac{1}{3 \epsilon} + \frac{194}{27}-\frac{7}{6} \gamma -
\frac{17}{36} \pi^2 - \frac{2}{3} \ln 2 - 2 \zeta(3) - \frac{1}{2} \ln
\frac{2 \mu}{H} + 
\nonumber \\ &\hspace{.0\textwidth} \bigg(\frac{2}{3\epsilon} +
\frac{127}{18} - 
\frac{17}{6} \gamma - \frac{1}{3} \pi^2 -\frac{4}{3} \ln 2-
\frac{3}{2} \ln 
\frac{2\mu}{H} \bigg) \ln \frac{\tau}{\tau_H} + \nonumber \\
& \hspace{.0\textwidth} \bigg(\frac{1}{2\epsilon} + \frac{8}{3} -
\gamma -\ln 2\bigg) \ln^2 \frac{\tau}{\tau_H} + \frac{1}{3} \ln^3
\frac{\tau}{\tau_H} + \mathcal{O}(\frac{\tau}{\tau_H}) +
\mathcal{O}(\epsilon) \bigg\} . \label{evad}
\end{align}
There is an equal contribution from the diagram with $\tau_1$ and
$\tau_2$ interchanged. Note that there is no dependence on $\ln k/\mu$
for $|k\tau| \ll 1$.

\subsection{Diagrams B and C}
The amputated versions of the diagrams with two external $G^R$
propagators are
\begin{equation}
\picc{
\DashLine(6,20)(22,20){2}
\CArc(32,20)(10,0,360)
\DashLine(42,20)(58,20){2}
\Text(66,36)[rt]{B}
\LongArrowArcn(32,20)(14,140,40)
\Text(32,35)[cb]{$\mathbf{p}'$}
\LongArrowArc(32,20)(14,220,320)
\Text(32,5)[ct]{$\mathbf{p}$}
\LongArrow(6,26)(18,26)
\Text(12,30)[cb]{$\mathbf{k}$}
\Text(2,16)[lt]{$\tau_1$}
\Text(60,16)[rt]{$\tau_2$}
} \hspace{.1\textwidth}
\picc{
\DashLine(6,20)(22,20){2}
\CArc(32,20)(10,90,270)
\DashCArc(32,20)(10,-90,90){2}
\DashLine(42,20)(58,20){2}
\Text(66,36)[rt]{C}
\LongArrowArcn(32,20)(14,140,40)
\Text(32,35)[cb]{$\mathbf{p}'$}
\LongArrowArc(32,20)(14,220,320)
\Text(32,5)[ct]{$\mathbf{p}$}
\LongArrow(6,26)(18,26)
\Text(12,30)[cb]{$\mathbf{k}$}
\Text(2,16)[lt]{$\tau_1$}
\Text(60,16)[rt]{$\tau_2$}
}.
\end{equation}
They translate to
\begin{align}
{\rm B}_{\rm amp}(k,\tau_1, \tau_2) &= \frac{(-i \lambda)^2}{2 H^8
  (\tau_1 \tau_2)^4} \int \frac{d^3 p d^3 
  p'}{(2 \pi)^3} \, \delta^3(\mathbf{k} -\mathbf{p} - \mathbf{p}')
  F(p', \tau_1, \tau_2) F(p, 
  \tau_1, \tau_2) \nonumber \\
& = \frac{-\lambda^2}{2 (2 \pi)^2 k H^8 (\tau_1 \tau_2)^4}
  \int_0^{\infty} dp \, p \int_{|p-k|}^{p+k} dp' \, p' \,
  F(p', \tau_1, \tau_2) F(p,\tau_1, \tau_2)
,  \label{genexprB} \\
{\rm C}_{\rm amp}(k,\tau_1,\tau_2) & = \frac{(-i)^2(-i \lambda)^2}{8
  H^8 (\tau_1 \tau_2)^4} \int \frac{d^3 p d^3 
  p'}{(2 \pi)^3} \, \delta^3(\mathbf{k} -\mathbf{p} - \mathbf{p}')
  G^R(p', \tau_1, \tau_2) G^R(p, 
  \tau_1, \tau_2) \nonumber \\
& = \frac{\lambda^2}{8 (2 \pi)^2 k H^8 (\tau_1 \tau_2)^4}
  \int_0^{\infty} dp \, p \int_{|p-k|}^{p+k} dp' \, p' \,
  G^R(p', \tau_1, \tau_2) G^R(p,\tau_1, \tau_2)
,  \label{genexprC}
\end{align}
where both diagrams have a factor $1/2$ for symmetry. Diagram C has an
additional factor $1/4$ from the vertex with three dashed lines
\eqref{qvertex}. We split the $p$ integral again in a small momentum
part and a large momentum part.

\subsubsection{Amputated diagrams for small internal momenta}
\label{smampbc}
For small internal momenta we use the expanded propagators
\eqref{expandedf} and \eqref{expandedgr}.

\paragraph{Diagram B.}
The integral is
\begin{multline}
\frac{-\lambda^2}{2(2 \pi)^2 k H^8 (\tau_1 \tau_2)^4}
\frac{H^4 \left(\tau_1 \tau_2 \right)^{2 \epsilon}}{4} \int_0^{M_{\rm cm}} dp
\int_{|p-k|}^{p+k} dp' \, \frac{(p p')^{2 \epsilon}}{(p p')^2} = \\
\frac{-\lambda^2 \left(\tau_1 \tau_2 \right)^{2\epsilon}}{8(2 \pi)^2 k
  H^4 (\tau_1 \tau_2)^4 (2 \epsilon -1)} \bigg(\int_0^k dp \, p^{-2+2 \epsilon}
\left((p+k)^{-1 +2 \epsilon} -(k-p)^{-1+2 \epsilon} \right) + \\ \int_k^{M_{\rm cm}} dp \,
p^{-2 +2 \epsilon} \left( (p+k)^{-1+2 \epsilon} - (p-k)^{-1+2 \epsilon} 
\right) \bigg). \label{diagblow}
\end{multline}
The integral on the middle line of \eqref{diagblow} is finite, but the
individual parts are infrared divergent. Therefore we calculate the
individual parts for $\epsilon > 1/2$, and in the end use analytic
continuation to $\epsilon \ll 1$. The integrals are (using $p=k x$)
\begin{align}
\int_0^1 dx \, x^{-2 + 2 \epsilon}\, (1+ x)^{-1 + 2 \epsilon} & =
\sum_{n=0}^{\infty} \frac{(-1)^n}{n!} \frac{\Gamma (1+n
  -2 \epsilon)}{\Gamma(1-2 \epsilon)} \, \int_0^1 dx \,x^{-2+n+2
  \epsilon} \nonumber \\ 
& = -\frac{1}{2 \epsilon} + \ln 2 + \mathcal{O}(\epsilon), \\
\int_0^1 dx \, x^{-2+2 \epsilon}\, (1-x)^{-1+2 \epsilon} & = B(-1 +
2 \epsilon, 2 \epsilon) = \frac{1}{\epsilon} -2 + \mathcal{O}(\epsilon), \\
\int_1^{{M_{\rm cm}}/k} dx \, x^{-2+2 \epsilon} \, (1+x)^{-1+2 \epsilon} & =
\int_{k/{M_{\rm cm}}}^1 dy \, \frac{y^{1-4 \epsilon}}{(1+y)^{1-2 \epsilon}}
\nonumber \\
&=1-\frac{k}{{M_{\rm cm}}} - \ln 2 + \ln\left(1+\frac{k}{{M_{\rm cm}}}\right) +
\mathcal{O}(\epsilon), \\
\int_1^{{M_{\rm cm}}/k} dx \, x^{-2+2\epsilon}\, (x-1)^{-1+2\epsilon} & =
\int_{k/{M_{\rm cm}}}^1 dy\, \frac{y^{1-4\epsilon}}{(1-y)^{1-2\epsilon}} =
\int_0^{1-k/{M_{\rm cm}}} dz \, z^{-1+2 \epsilon}\,(1-z) + \mathcal{O}(\epsilon)
\nonumber \\ 
&= \frac{1}{2\epsilon} -1 + \frac{k}{{M_{\rm cm}}} + \ln\left(1-\frac{k}{{M_{\rm cm}}}
\right) + \mathcal{O}(\epsilon),
\end{align}
where we have used analytic continuation in the first two integrals
and $y=1/x$ and $z=1-y$ in the latter two. The right hand side of
equation \eqref{diagblow} becomes
\begin{multline}
\frac{-\lambda^2 (k^2 \tau_1\tau_2)^{2 \epsilon}}{8 (2 \pi)^2 k^3 H^4
  (\tau_1 \tau_2)^4 (2 \epsilon -1)} \bigg(\frac{-2}{\epsilon} + 4 - 2
  \frac{k}{{M_{\rm cm}}} + \ln \frac{1+k/{M_{\rm cm}}}{1-k/{M_{\rm cm}}} + \mathcal{O}(\epsilon) \bigg)
=  \\
\frac{-\lambda^2}{4(2\pi)^2 k^3 H^4 (\tau_1 \tau_2)^4}
  \bigg(\frac{1}{\epsilon} + \frac{k}{{M_{\rm cm}}} -\frac{1}{2} \ln
  \frac{{M_{\rm cm}}+k}{{M_{\rm cm}}-k} +2 \,\ln(k^2 \tau_1\tau_2) +
  \mathcal{O}(\epsilon) \bigg) . \label{resdiagblow}
\end{multline}

\paragraph{Diagram C.}
From equations \eqref{genexprC} and \eqref{expandedgr} we see directly
that diagram C does not give late time contributions and also does not
have an infrared divergence.

\subsubsection{Amputated diagrams for large internal momenta}
The contributions from large internal momenta can be calculated in a
similar way as is used for diagram A in section
\ref{largemomentadiaga}. 

\paragraph{Diagram B.}
The sum of integrals
\begin{equation}
\sum_{i=1}^3 \sum_{j=1}^3 \int_{M_{\rm cm}}^{\Lambda a(\tau_{\beta})} dp\, p \,
F_i(p,\tau_1, \tau_2) 
\int_{p-k}^{p+k} dp' \, p' \,F_j(p',\tau_1, \tau_2),
\end{equation}
(where $\tau_{\beta}=\tau_1,\tau_2$, depending on which time is
earlier), is equal to
\begin{align}
& \bigg[\frac{\cos p \Delta \tau}{p} \left(\frac{\cos(p+k)\Delta
    \tau}{p+k} - \frac{\cos(p-k)\Delta \tau}{p-k} \right) + \nonumber \\
& \hspace{.05\textwidth} - \tau_1 \tau_2 \sin k \Delta \tau
 \bigg({\rm Si}(2 p \Delta \tau) +2 \frac{\cos^2 p \Delta \tau}{p
    \Delta \tau} \bigg) + \frac{(\tau_1 \tau_2)^2}{\Delta \tau} \sin
    k \Delta \tau \bigg(p+ \frac{\sin 2 p \Delta \tau}{2 \Delta \tau}
    \bigg) \bigg]^{\Lambda a(\tau_{\beta})}_{M_{\rm cm}} \nonumber + \\
&\int_{M_{\rm cm}}^{\Lambda a(\tau_{\beta})} dp\, \frac{\cos p \Delta
    \tau}{p} \bigg\{ 
    \frac{\cos (p+k)\Delta \tau}{(p+k)^2} - \frac{\cos (p-k) \Delta
    \tau}{ (p-k)^2} +\nonumber \\
&\hspace{.05\textwidth} \Delta \tau \bigg( \frac{\sin(p+k) \Delta
    \tau}{p+k} - \frac{\sin(p-k) \Delta\tau}{p-k} \bigg) -p \tau_1
    \tau_2 \bigg(\frac{\cos(p+k)\Delta \tau}{p+k} - \frac{\cos (p-k)
    \Delta \tau}{p-k} \bigg) \bigg\} . \label{verylongeq2}
\end{align}
The only ultraviolet term comes from the last term of the boundary
term and is, including the correct prefactor:
\begin{equation}
\frac{-\lambda^2 \, \sin k \Delta \tau}{8 (2\pi)^2 k H^4 (\tau_1
  \tau_2)^2 \Delta \tau} \bigg[ p + \frac{\sin 2 p \Delta \tau}{2
  \Delta \tau} \bigg]^{\Lambda a(\tau_{\beta})}_{M_{\rm cm}} . \label{uvdiagb}
\end{equation}
The only term that gives late time contributions is the first line in
the integral. It is
\begin{equation}
\frac{-\lambda^2}{4(2\pi)^2 k^3 H^4 (\tau_1 \tau_2)^4} \bigg( -
\frac{k}{{M_{\rm cm}}} + \frac{1}{2} \ln \frac{{M_{\rm cm}}+k}{{M_{\rm cm}}-k} +
\mathcal{O}(\tau_i^2)  \bigg) . \label{ltclmb}
\end{equation}

\paragraph{Diagram C.}
The sum of integrals
\begin{equation}
\sum_{i=1}^3 \sum_{j=1}^3 \int_{M_{\rm cm}}^{\Lambda a(\tau_{\beta})} dp\, p \,
G^R_i(p,\tau_1, \tau_2) 
\int_{p-k}^{p+k} dp' \, p' \,G^R_j(p',\tau_1, \tau_2),
\end{equation}
is equal to
\begin{align}
& \bigg[\frac{\sin p \Delta \tau}{p} \left(\frac{\sin(p+k)\Delta
    \tau}{p+k} - \frac{\sin(p-k)\Delta \tau}{p-k} \right) + \nonumber \\
& \hspace{.05\textwidth} + \tau_1 \tau_2 \sin k \Delta \tau
 \bigg(3\, {\rm Si}(2 p \Delta \tau) -2 \frac{\sin^2 p \Delta \tau}{p
    \Delta \tau} \bigg) + \frac{(\tau_1 \tau_2)^2}{\Delta \tau} \sin
    k \Delta \tau \bigg(p- \frac{\sin 2 p \Delta \tau}{2 \Delta \tau}
    \bigg) \bigg]^{\Lambda a(\tau_{\beta})}_{M_{\rm cm}} \nonumber + \\
&\int_{M_{\rm cm}}^{\Lambda a(\tau_{\beta})} dp\, \frac{\sin p \Delta
    \tau}{p} \bigg\{ 
    \frac{\sin (p+k)\Delta \tau}{(p+k)^2} - \frac{\sin (p-k) \Delta
    \tau}{ (p-k)^2} +\nonumber \\
&\hspace{.05\textwidth} -\Delta \tau \bigg( \frac{\cos(p+k) \Delta
    \tau}{p+k} - \frac{\cos(p-k) \Delta\tau}{p-k} \bigg) -p \tau_1
    \tau_2 \bigg(\frac{\sin(p+k)\Delta \tau}{p+k} - \frac{\sin (p-k)
    \Delta \tau}{p-k} \bigg) \bigg\} .
\end{align}
Only the last term of the boundary term is ultraviolet divergent:
\begin{equation}
\frac{\lambda^2 \, \theta(\tau_1 - \tau_2) \, \sin k \Delta \tau}{8
  (2\pi)^2 k H^4 (\tau_1 
  \tau_2)^2 \Delta \tau} \bigg[ p - \frac{\sin 2 p \Delta \tau}{2
  \Delta \tau} \bigg]^{\Lambda a(\tau_{\beta})}_{M_{\rm cm}} . \label{uvdiagc}
\end{equation}
The diagram with the vertices exchanged gives the same result, except
that $\theta(\tau_1-\tau_2)$ is replaced by
$\theta(\tau_2-\tau_1)$. There are no further late time contributions.

\paragraph{Ultraviolet divergences.}
The ultraviolet divergent terms of diagrams B \eqref{uvdiagb}, C
\eqref{uvdiagc}, and C with the vertices exchanged, add up to
\begin{equation}
\bigg[\frac{-\lambda^2 \, \sin k \Delta \tau}{8(2 \pi)^2 k H^4 (\tau_1
    \tau_2)^2 \Delta \tau} \frac{\sin 2 p \Delta \tau}{\Delta\tau}
    \bigg]^{\Lambda a(\tau_{\beta})}_{M_{\rm cm}} ,
\end{equation}
which is finite and does not give late time contributions. 

\subsubsection{Attaching the external lines}
Adding the small and large momenta contributions, we obtain for the
amputated diagrams B and C:
\begin{equation}
{\rm B}_{\rm amp}(k,\tau_1, \tau_2) + {\rm C}_{\rm amp}(k,
\tau_1, \tau_2) = \frac{-\lambda^2}{4 (2\pi)^2 k^3 H^4 (\tau_1
  \tau_2)^4} \bigg(\frac{1}{\epsilon} + 2 \ln(k^2 \tau_1 \tau_2) +
\mathcal{O}(\tau_i) +\mathcal{O} (\epsilon) \bigg),
\end{equation}
where the dependence on ${M_{\rm cm}}$ has dropped out. The full correlation
function is obtained by
\begin{equation}
- \int_{\tau_H}^{\tau} d\tau_1 \int_{\tau_H}^{\tau} d\tau_2 \,
G^R(k,\tau,\tau_1) G^R(k,\tau,\tau_2) \Big( {\rm B}_{\rm
  amp}(k,\tau_1,\tau_2) + {\rm C}_{\rm
  amp}(k,\tau_1,\tau_2) \Big).
\end{equation}
Because the external momentum $k$ is small, i.e. $|k\tau_i| \ll 1$, we
can use the expanded version of the $G^R$ propagator \eqref{expangr}
(or the one of \eqref{expandedgr}, but this gives only corrections of
order $\mathcal{O} (\epsilon)$). Using the integrals
\begin{align}
& \int_{\tau_H}^{\tau}d\tau_1 \int_{\tau_H}^{\tau} d\tau_2\,
\frac{(\tau^3-\tau_1^3)(\tau^3- \tau_2^3)}{(\tau_1 \tau_2)^4} =
\frac{1}{9} + \frac{2}{3} \ln \frac{\tau}{\tau_H} + \ln^2
\frac{\tau}{\tau_H} + \mathcal{O}(\frac{\tau}{\tau_H}), \\
& \int_{\tau_H}^{\tau}d\tau_1 \int_{\tau_H}^{\tau} d\tau_2\,
\frac{(\tau^3-\tau_1^3)(\tau^3- \tau_2^3)}{(\tau_1 \tau_2)^4} \, \ln(
k^2 \tau_1 \tau_2) = \frac{1}{27} \bigg(2 + 6 \ln(- k \tau_H) + \nonumber
\\ & \hspace{.06\textwidth} 12 \Big(1 +3 \ln(- k \tau_H )\Big) \ln
\frac{\tau}{\tau_H} 
+ 27 \Big(1+2 \ln (-k \tau_H) \Big)  \ln^2 \frac{\tau}{\tau_H} + 27 \ln^3
\frac{\tau}{\tau_H} \bigg) +  \mathcal{O}(\frac{\tau}{\tau_H}),
\end{align}
this becomes
\begin{multline}
\frac{\lambda^2}{36 (2\pi)^2 k^3} \bigg( \frac{1}{9 \epsilon} +
\frac{4}{27} + \frac{4}{9} \ln (-k\tau_H) + \Big( \frac{2}{3\epsilon} +
\frac{8}{9} + \frac{8}{3} \ln (-k \tau_H) \Big) \ln \frac{\tau}{\tau_H}
+ \\ \Big(\frac{1}{\epsilon}+2 + 4 \ln(- k \tau_H) \Big) \ln^2
\frac{\tau}{\tau_H} + 2 \ln^3 \frac{\tau}{\tau_H} + \mathcal{O}
(\frac{\tau}{\tau_H} ) + \mathcal{O}(\epsilon) \bigg). \label{evbc}
\end{multline}

\bibliography{refs}

\end{document}